\documentclass[amsmath,amssymb,nofootinbib,notitlepage,superscriptaddress,twocolumn]{revtex4-1}
\pdfoutput=1

\usepackage{aas_macros,graphicx}
\usepackage[bottom]{footmisc}
\usepackage[colorlinks=true]{hyperref}
\let\originalleft\left
\let\originalright\right
\renewcommand{\left}{\mathopen{}\mathclose\bgroup\originalleft}
\renewcommand{\right}{\aftergroup\egroup\originalright}

\newcommand{\ab}[1]{\left|#1\right|}
\newcommand{\br}[1]{\left[#1\right]}
\newcommand{\cu}[1]{\left\{#1\right\}}
\newcommand{\pa}[1]{\left(#1\right)}

\newcommand{\ed}{\mathop{}\!\mathrm{d}}
\newcommand{\pd}{\mathop{}\!\partial}

\DeclareMathOperator\arcsinh{arcsinh}
\DeclareMathOperator\sign{sign}
\DeclareMathOperator\sn{sn}

\begin{document}

\title{The Shape of the Black Hole Photon Ring:\\
A Precise Test of Strong-Field General Relativity}

\author{Samuel E. Gralla}
\email{sgralla@email.arizona.edu}
\affiliation{Department of Physics, University of Arizona, Tucson, AZ 85721, USA}
\author{Alexandru Lupsasca}
\email{lupsasca@princeton.edu}
\affiliation{Princeton Gravity Initiative, Princeton University, Princeton, NJ 08544, USA}
\affiliation{Society of Fellows, Harvard University, Cambridge, MA 02138, USA}
\author{Daniel P. Marrone}
\email{dmarrone@email.arizona.edu}
\affiliation{Steward Observatory, University of Arizona, Tucson, AZ 85721, USA}

\begin{abstract}
We propose a new test of strong-field general relativity (GR) based on the universal interferometric signature of the black hole photon ring.  The photon ring is a narrow ring-shaped feature, predicted by GR but not yet observed, that appears on images of sources near a black hole.  It is caused by extreme bending of light within a few Schwarzschild radii of the event horizon and provides a direct probe of the unstable bound photon orbits of the Kerr geometry.  We show that the precise shape of the observable photon ring is remarkably insensitive to the astronomical source profile and can therefore be used as a stringent test of GR.  We forecast that a tailored space-based interferometry experiment targeting M87* could test the Kerr nature of the source to the sub-sub-percent level.
\end{abstract}

\maketitle

\section{Introduction}

\begin{figure*}
	\centering
	\includegraphics[width=\textwidth]{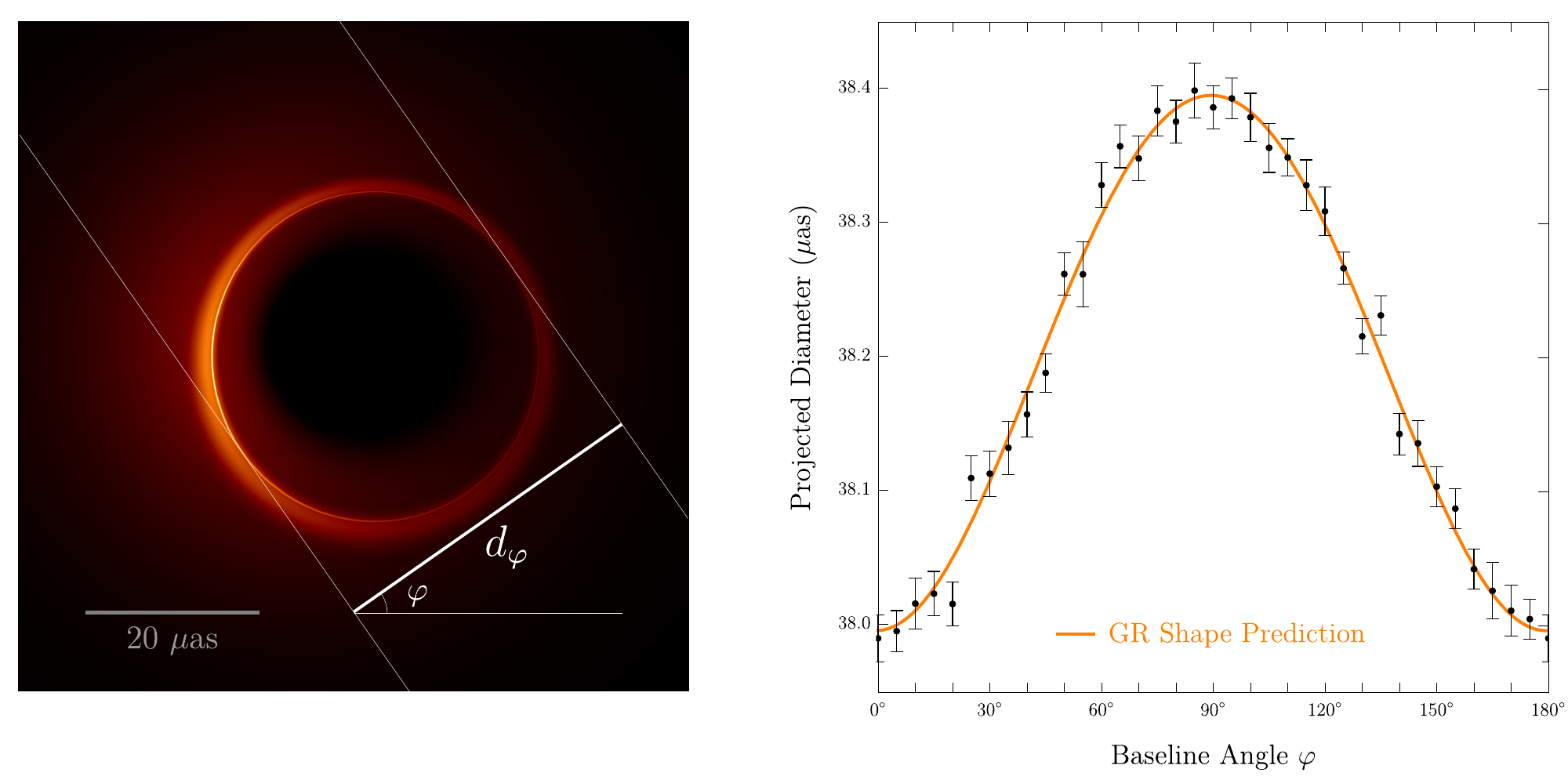}
	\caption{A proposed test of the Kerr metric and its strong-field gravitational lensing.  General Relativity (GR) predicts that light orbiting near the event horizon of a black hole produces a sequence of narrow ``photon rings'' on images of the surrounding emission.  On the left, we show a model for the time-averaged observational appearance of M87*, resolving the main image of the accretion disk as well as the first two of the infinitely many photon rings.  We propose a space interferometry experiment to measure the projected diameter $d_\varphi$ of the second photon ring at different angles $\varphi$.  On the right, we show the forecasted experimental result, compared with the best-fit GR prediction for $d_\varphi$.  This proposed experiment will test the GR prediction for the photon ring shape with a precision of 0.04\% [Eq.~\eqref{eq:TestGR}].}
	\label{fig:PhotonRingExperiment}
\end{figure*}

General relativity (GR) is a pillar of modern physics and a workhorse of astronomical modeling.  As the theory lacks any intrinsic scale, its predictions are organized by a dimensionless ratio of system parameters.  If $M$ is a typical mass scale and $R$ is a typical distance scale, we may construct the dimensionless quantity
\begin{align}
	\Phi=\frac{GM}{Rc^2},
\end{align}
where $G$ is Newton's constant and $c$ is the speed of light.  In the Newtonian limit, this quantity is just (minus) the gravitational potential, and more generally we may take it to represent the strength of the gravitational field.

GR has been tested with exquisite accuracy in the regime $\Phi\ll1$ of weak gravitational fields \cite{Berti2015,Will2014,LIGO2017}.  The strong-field regime is comparably lacking in precise tests, with two significant exceptions.  First, binary pulsar measurements \cite{Kramer2013} have sensitively tested the GR prediction that the motion of a body depends only on its total mass and not on its internal composition (the ``strong equivalence principle'').  Second, the spectacularly bright first gravitational-wave source GW150917 \cite{LIGO2016a} has been used for a few-percent-level test of the GR prediction for the gravitational waves generated by a black hole merger \cite{LIGO2016b}.

While very impressive, these tests cover only a small fraction of the important strong-field phenomenology of GR.  In particular, there is as yet no precise, direct test of the prediction that black holes are described by the Kerr metric.  The Kerr assumption underlies an enormity of important astronomical modeling, and paradoxes raised by black holes play a driving role in theoretical physics.  We therefore consider it imperative to seek direct tests of the Kerr black hole prediction.  More generally, the foundational status of GR as our most fundamental description of gravity demands thorough testing of the theory in all its regimes.

In this paper, we will discuss a new potential test of strong-field GR based on the observation of a characteristic ``photon ring'' predicted to arise due to black hole lensing.  Such a ring first appeared in a simulated image produced by Luminet in 1979 \cite{Luminet1979}, although some elements of the underlying theory date back to Bardeen in 1973 \cite{Bardeen1973}, to Darwin in 1959 \cite{Darwin1959}, and even to Hilbert in 1917 \cite{Hilbert1917}.  The essential realization is that since light can orbit (unstably) around black holes, it follows that for any source emitting in all directions, and for any detector collecting light, some fraction of emitted photons loop around the black hole before reaching the detector.  An even smaller fraction will loop an additional time, and so on.  In principle, therefore, a camera aimed at a black hole sees every object in the universe, each appearing an infinite number of times!  The orbiting photons all arrive near a closed, ``critical curve'' on the image plane, creating a ring-like feature known as the photon ring \cite{Bardeen1973,Beckwith2005,Falcke2000,Gralla2019,Johnson2020,GrallaLupsasca2020a}.\footnote{In this paper, we use ``photon ring'' to mean any observable ring-like feature associated with orbiting photons.  The models we consider herein exhibit multiple photon rings, one for each image of the disk.}

The last fifteen months have witnessed several rapid developments pushing the photon ring from a theoretical concept towards an observational reality.  First, the Event Horizon Telescope (EHT) collaboration released interferometric observations of the black hole at the center of the galaxy M87 (henceforth M87*) \cite{EHT2019a,EHT2019b,EHT2019c,EHT2019d,EHT2019e,EHT2019f}.  These provide new detail about the emission structure on horizon scales and, as we shall see, help demonstrate the suitability of the source for future photon ring observations.

Next came a succession of theoretical developments.  Ref.~\cite{Gralla2019} showed that the photon ring from sources like M87*, previously thought of as a single continuous entity, should in fact consist of a sequence of individual rings converging to the critical curve.  Ref.~\cite{Johnson2020} then noted that these rings are \textit{persistent sharp features}, and hence should dominate time-averaged interferometric observations on suitably long baselines.  In particular, a ring of diameter $d$ and width $w$ dominates the interferometric signal in the range 
\begin{align}
	\label{eq:UniversalRegime}
	1/d\ll u\ll 1/w,
\end{align}
and its signature is a visibility (Fourier transform of the image intensity) that oscillates along a radial baseline $u$, with a periodicity related to the ring diameter.   

Ref.~\cite{Gralla2020} provided complete details of this universal observational signature, calculating the visibility in the regime \eqref{eq:UniversalRegime} of a narrow feature tracing an arbitrary plane curve.  For a closed, convex curve, the signature is
\begin{align}
	\label{eq:UniversalSignature}
	V&=\frac{e^{-2\pi iC_\varphi u}}{\sqrt{u}}\pa{\alpha_\varphi^Le^{-\frac{i\pi}{4}}e^{i\pi d_\varphi u}+\alpha_\varphi^Re^{\frac{i\pi}{4}}e^{-i\pi d_\varphi u}},
\end{align}
where $(u,\varphi)$ are polar coordinates on the visibility plane.  The functions $d_\varphi$ and $C_\varphi$ are the angle-dependent projected diameter and centroid, respectively.  We illustrate $d_\varphi$ in Fig.~\ref{fig:PhotonRingExperiment}---see Refs.~\cite{Gralla2020,GrallaLupsasca2020c} for precise definitions of $d_\varphi$ and $C_\varphi$.  These functions encode the shape of the curve, while $\alpha_\varphi^L$ and $\alpha_\varphi^R$ encode its intensity profile.  Ref.~\cite{GrallaLupsasca2020c} presented an explicit formula for reconstructing the full curve from the interferometric data $C_\varphi$ and $d_\varphi$.

While the full complex visibility \eqref{eq:UniversalSignature} encodes both $d_\varphi$ and $C_\varphi$, the visibility \textit{amplitude} depends on $d_\varphi$ only:
\begin{align}
	\label{eq:VisibilityAmplitude}
	\ab{V}=\frac{1}{\sqrt{u}}\sqrt{\pa{\alpha_\varphi^L}^2+\pa{\alpha_\varphi^R}^2+2\alpha_\varphi^L\alpha_\varphi^R\sin\pa{2\pi d_\varphi u}}.
\end{align}
This function oscillates with period $1/d_\varphi$ between maximum and minimum values set by the sum and difference of $\alpha_L$ and $\alpha_R$.  Thus, one can infer the projected diameter from the visibility amplitude alone by measuring its oscillation frequency on long baselines.  Ref.~\cite{GrallaLupsasca2020c} explored the shape information available from just the set of projected diameters $d_\varphi$, without the centroids $C_\varphi$.

These recent results suggest a conceptually simple yet powerful new test of strong-field GR: measure the shape of a photon ring and compare it to the GR prediction.  But is such a measurement feasible, and would it even really test GR?

Using a suite of reasonable M87* source models informed by the EHT observations, we show that the photon ring shape, as measured via its interferometric signature, is an astonishingly precise prediction of GR, independent of astrophysical details.  This theoretical result establishes exceptional promise for shape measurements as tests of GR.  However, our modeling also reveals that the photon ring signal is expected to be rather weak, with long-baseline visibility on the order of a few tenths of a milliJansky (mJy).  As such, there will be significant experimental challenges involved in making such a measurement, which can only be carried out with a space interferometer.  However, we show that if these challenges are met and the required sensitivity achieved, then shape measurements with sub-sub-percent precision are possible.  Such measurements would provide a new test of strong-field GR with unprecedented precision, probing for the first time the detailed geometry of the Kerr black hole.  The measurement concept and forecast are illustrated in Figs.~\ref{fig:PhotonRingExperiment} and \ref{fig:Orbit}.

This paper is organized as follows.  In Sec.~\ref{sec:Overview}, we give an overview of our reasoning and results, which we then describe in more detail in Secs.~\ref{sec:Models} and \ref{sec:Forecast}.  We then compare with other related work in Sec.~\ref{sec:Comparison}, before concluding with our outlook in Sec.~\ref{sec:Summary}.

\section{Overview}
\label{sec:Overview}

As described in the introduction, recent observational and theoretical progress has set the stage for the photon ring shape to be used as a precise test of GR.  However, standing between these recent results and a compelling GR test are three important questions:
\begin{enumerate}
	\item The approximation \eqref{eq:UniversalSignature} becomes exact as the ring width goes to zero and the range of validity \eqref{eq:UniversalRegime} becomes infinite.  For the photon rings of small-but-finite width visible in images of reasonable source models, does the analytic signature \eqref{eq:UniversalSignature} really dominate in some finite range of baselines, with sufficient fidelity to infer the photon ring shape $d_\varphi$ and $C_\varphi$?  In other words: Assuming GR is correct, is it possible  \textit{even in principle} (with a perfect detector) to measure the shape of a photon ring?
	\item Does the precise shape of the photon ring depend sensitively on the astrophysical source profile, or is it a universal (matter-independent) prediction of GR?\footnote{The shape of the critical curve follows entirely from GR, but is not in itself observable.  Successive photon rings converge to the critical curve, with deviations exponentially suppressed in orbit number $n$ but still important at small $n$ (see Sec.~\ref{sec:CriticalCurve} and Fig.~\ref{fig:PhotonRings}).}  In other words, to what extent does measuring the shape of the photon ring really test GR?
	\item Can this test be realized at a precision that provides an interesting test of GR?
\end{enumerate}

We are pleased to report encouraging results on all three fronts, as described in the following three sections.

\subsection{The photon ring shape is measurable in principle}

Our focus is on the putative black hole M87*, which is believed to possess a geometrically thick accretion disk.  Although the Kerr spacetime is stationary, axisymmetric, and reflection-symmetric about the equatorial plane, the necessarily turbulent flow of matter accreting onto a black hole \cite{Balbus1991} will undoubtedly break these symmetries on short timescales.  However, absent some persistent external boundary conditions,\footnote{One example of a symmetry-breaking boundary condition is given by an accretion disk that is misaligned with the black hole spin.  In this setting, we would still expect a regular structure to emerge after time-averaging, which could be modeled by a stationary (but not axisymmetric) phenomenological model.  The observational appearance of tilted disk configurations has been studied in Ref.~\cite{White2020}, albeit not with the resolution required to see the emergence of the universal signature on long baselines.} one expects the symmetries of the underlying Kerr spacetime to reemerge after time-averaging over some suitable timescale.  The timescale for M87* must lie somewhere between its light-crossing time (a few days) and the orbital period of nearby matter ($\sim$1-2 months), and simulations generally show non-axisymmetric structures forming and disappearing over a period of a week or two.  The photon ring is present in every snapshot simulated image, but is most prominent in the time-averaged image \cite{Johnson2020}, whose Fourier transform is the time-averaged radio visibility.

To study the time-averaged observational appearance of M87*, we will consider models that share the symmetries of the underlying Kerr spacetime.  We adopt an ``equatorial approximation'' in which the source is assigned a stationary, axisymmetric intensity profile in the plane perpendicular to the black hole spin (representing emission from a disk-like source), augmented with a fudge factor to account for the effects of geometrical thickness.  This approach is capable of reproducing the time-averaged observational appearance of simulation-based models \cite{EHT2019e}, while also facilitating access to a much larger range of reasonable source profiles.  We mainly confine our attention to models that produce visibility amplitudes roughly consistent with the M87* observations.  This limits the observer inclination $\theta_o$ from the black hole spin axis to less than $\sim$30$^\circ$ and constrains the mass-to-distance ratio $M/D$ within a factor of $\sim$2-3, while still allowing the black hole spin $a$ to take any value $0<a<M$.

We analyze this class of models in detail.  We resolve the first ($n=1$) and second ($n=2$) of the infinitely many rings converging to the critical curve.\footnote{The index $n$ refers to the approximate number of half-orbits executed by photons before reaching the detector \cite{Gralla2019,Johnson2020,GrallaLupsasca2020a,Himwich2020}.  In the nomenclature of Ref.~\cite{Gralla2019}, our $n=1$ photon ring is the ``lensing ring'', while our $n\ge2$ rings comprise the photon ring.  In the nomenclature of Refs.~\cite{Johnson2020,Himwich2020}, the rings labeled by $n$ are ``subrings'', with the ``photon ring'' denoting the entire set.}  The first ring is usually too wide to admit the separation of scales required for the existence of a universal regime \eqref{eq:UniversalRegime}, but the second ring cleanly presents the signature \eqref{eq:UniversalSignature} in all cases.  To quantify this statement, we adopt a canonical range of baseline lengths $u\sim285$-315~G$\lambda$ in which we fit for the parameters $\cu{d_\varphi,C_\varphi,\alpha_L^\varphi,\alpha_R^\varphi}$ at every angle $\varphi$.  In all models, we find excellent fits as judged by normalized residuals.  We take special care in judging the accuracy of the fit for $d_\varphi$ (e.g., Fig.~\ref{fig:ProjectedDiameter}), since it will be the focus of our experimental proposal.  This analysis establishes that a shape measurement is possible in principle for this class of models.

Since these models cover a very large range of potential time-averaged observational appearances of M87*, and since for every model there is a clean, measurable photon ring signature \eqref{eq:ComplexVisibility} over some range of baselines $u$, our answer to question 1 is: \textit{Yes, the measurement is possible in principle.}

\subsection{The photon ring shape is predicted by GR}

We have argued that the shape of the ($n=2$) photon ring from M87* is in principle observable via its interferometric signature.  To determine the extent to which such a measurement tests GR, we need a GR prediction for this shape.  However, GR only predicts the shape of the critical curve, which is not directly observable.  On the other hand, the shape of the photon ring, which is measurable, depends at least somewhat on the astrophysical source profile, to a degree which has not yet been quantified.  For the expected emission profiles, photon rings are typically offset outwards from the critical curve.  At the level of precision that we envision, we require a GR prediction for \textit{photon rings}, not the critical curve.

Our study of the models described above provides just such a prediction.  We offer two levels of precision.  At the percent level, the answer is clear: the photon ring is an ellipse.  That is, we fit the inferred $d_\varphi$ and $C_\varphi$ to the functional forms of an ellipse [Eq.~(39) of Ref.~\cite{GrallaLupsasca2020c} together with the freedom to shift and rotate], finding that, for all source models, the best-fit ellipse has percent-level errors.  This implies a percent-level limit for GR tests via the photon ring shape.  With more careful analysis, this limit can undoubtedly be improved as well as extended to higher observer inclinations, where a non-elliptical shape is expected.  In such a line of development, one would naturally use a larger family of shapes, such as the lima\c{c}on \cite{Farah2020} or phoval \cite{GrallaLupsasca2020c}.

If only the visibility amplitude is measured (and not the phase), then only $d_\varphi$ can be measured (and not $C_\varphi$).  However, for $d_\varphi$ alone, we find a remarkable improvement in the precision of the GR prediction.  Whereas the ellipse model discussed above works to the percent level, the addition of a single parameter improves this to at least a part in $10^5$!  This parameter corresponds to a certain sum of an ellipse and a circle, producing a ``circlipse'' \cite{GrallaLupsasca2020c}.  To wit, the model is
\begin{align}
	\label{eq:Circlipse}
	d_\varphi=R_0+\sqrt{R_1^2\cos^2\pa{\varphi-\varphi_0}+R_2^2\sin^2\pa{\varphi-\varphi_0}}.
\end{align}
Of course, one can always make a fit work arbitrarily well by adding more parameters.  We are satisfied with the circlipse because it contains the same number of physical parameters as the GR prediction for the critical curve (namely, three). The critical curve depends on the black hole mass $M$, spin $a$ and observer inclination $\theta_o$, whereas the circlipse depends on the three radii $R_0$, $R_1$, and $R_2$.  (The offset angle $\varphi_0$ is degenerate with the orientation of the camera.)  The functional form \eqref{eq:Circlipse} holds across the wide range of disk-like, modestly inclined models that we consider, and hence can be considered a bona fide GR prediction in that regime, independent of astrophysical details.

Measuring $d_\varphi$ fixes the ``hull'' of the photon ring, but not its complete shape \cite{GrallaLupsasca2020c}.  We therefore predict that the hull of the photon ring is a circlipse.  While testing this prediction is perhaps viscerally less appealing than a full shape measurement, at core it is no less compelling: Here we have a precise functional form \eqref{eq:Circlipse} predicted by strong-field GR to appear, in a nearby source, to at least a part in $10^5$.  One can hardly hope for a more stringent test of the theory.  Our answer to question 2 is: \textit{The measurement can test GR with extraordinarily high precision.}

\subsection{The photon ring shape is measurable in practice}
\label{sec:Measurable}

Can this measurement be made in practice?  A mission targeting M87* must meet the following requirements:
\begin{enumerate}
	\item It must operate at frequencies where the source is optically thin.  The EHT results suggest that a good target observation frequency is $\nu\gtrsim230$~GHz.
	\item It must reach baselines in the universal regime \eqref{eq:UniversalRegime}.  Our modeling indicates that a good target baseline length is $u\sim300$~G$\lambda$.
	\item It must be sensitive enough to detect the universal signature.  Our modeling indicates that a good target sensitivity (standard deviation of the complex thermal noise) is $\sigma\lesssim0.1$-0.2~mJy within the coherent integration period.
	\item It must sample the Fourier plane finely enough to tease out the periodicity in the universal signature and thereby infer the photon ring shape.  The expected ring size suggests that a good target baseline density is $\Delta u\lesssim1$~G$\lambda$.
	\item It must be able to observe the same portion of the Fourier plane repeatedly, so as to average out any non-universal fluctuations.  The expected variability of M87* suggests that observations should be separated by weeks or months.
	\item It must be able to probe at least a few different angles $\varphi$ (ideally very many), such that shape tests are possible.
\end{enumerate}

To meet these requirements, we envision a two-element interferometer with a ``primary'' station in a distant Earth orbit, and a ``secondary'' station either on Earth or in low-Earth orbit.  Observations would be conducted at or above 230~GHz to ensure that the source is optically thin (item 1).  The wide primary orbit would place the typical baseline in the universal regime (item 2).  In this paper, we argue that the desired sensitivity and sampling (items 3 and 4) can be achieved via large collecting area, sensitive coherent receivers, and wide instantaneous bandwidths, with orbital determinations and initial fringe-finding achievable using careful experimental techniques.  The baseline will sample each $d_\varphi$ twice per orbit (angles $\varphi$ and $\varphi+\pi$ correspond to the same physical $d_\varphi$), enabling averaging over source fluctuations (item 5) at any angle $\varphi$ (item 6).  Since absolute phase measurements at the precision required for long-time averaging are technically unrealistic at present, we consider only incoherent averaging, i.e., we restrict consideration to the visibility amplitude only.  

\begin{figure}
	\centering
	\includegraphics[width=\columnwidth]{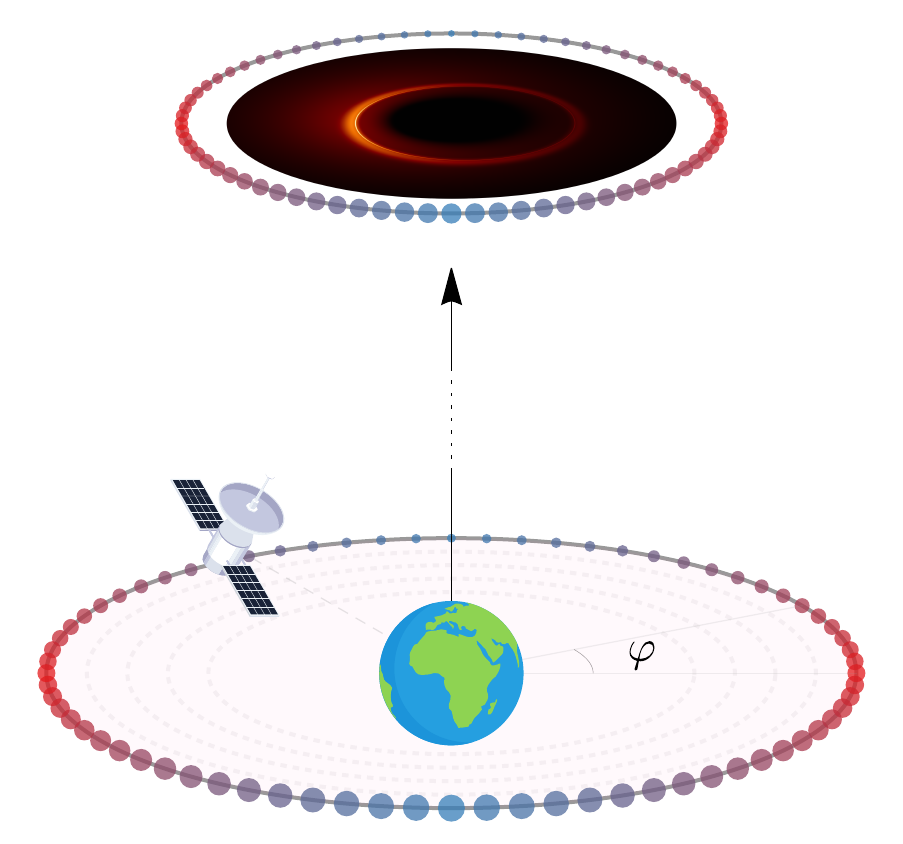}
 	\caption{Illustration of the configuration used for the experimental forecast.  A satellite orbits the Earth in the plane perpendicular to the line of sight to the target M87*, observing the source at various angles around the orbit.  Shape measurements are already possible with just a handful of observations, but in this paper we consider a scenario in which data points are collected from every 5$^\circ$ around the orbit.  (This would require multiple orbits if using a ground station.)  These observations are depicted with a colored dot, whose coloration matches that in Fig.~\ref{fig:MCMC}\textcolor{red}{(e)} below.}
	\label{fig:Orbit}
\end{figure}

We analyze in detail a fiducial configuration (Fig.~\ref{fig:Orbit}) where the primary orbits with a 720-hour ($\sim$1 month) period in the plane perpendicular to M87*,\footnote{In reality, such an orbit will be pertubed by the Moon, but the precise details of the orbit are not critical for this investigation of the potential experimental sensitivity.} and the secondary is a ground station. We imagine conducting ``observing runs'' of 2 hours, during which time the primary sweeps out 1 degree of its orbit.  We suppose that a coherent integration time of 5 minutes can be sustained, yielding 24 integration periods per run. Each observation is assumed to achieve a sensitivity of 0.14~mJy in each of 32 1-GHz bands surrounding 230~GHz, subdivided from a full-band fringe search.  This results in $24\times32=768$ complex visibilities per 2-hour run.  These fill up the range $u\sim285$-330~G$\lambda$, covering 8-9 oscillations of the expected universal signature \eqref{eq:UniversalSignature}.

The target of $\sim$0.1-0.2~mJy sensitivity for the single baseline of this experiment merits some discussion.  The flux density estimates of Sec.~\ref{sec:Models} indicate that sensitivity at this order of magnitude is critical.  For a bandwidth of 1~GHz, set by the need to clearly resolve the oscillations in visibility amplitude, and an integration time of 5 minutes, the required baseline system-equivalent flux density (SEFD) is approximately 100~Jy.  For reference, the median ALMA SEFD for the EHT M87* observations was 74~Jy.  A space-ground experiment that combines the existing ALMA and a large inflatable reflector \cite{Walker2019} illuminating a superconducting parametric amplifier \cite{HoEom2012} can meet the sensitivity target, so such an experiment is foreseeable with technologies that are approaching viability.

A realistic instrument will face many other challenges associated with the Very-Long-Baseline Interferometry (VLBI) measurement.  The proposed integration time is long compared to atmospheric timescales and pushes the limits of hydrogen masers, which may necessitate a space-space scheme in which a common reference signal is shared between satellites.  Fringe-finding in the absence of significant compact flux will require unconventional fringe tracking methods (e.g., between simultaneously observed frequencies) or hybrid constellations that provide short reference baselines.  Orbital determination, or at least the precise knowledge of the derivatives of the orbital separation, will also be essential to extended coherent integrations and may require optical ranging between satellites.  Other investigations of millimeterwave space VLBI have considered other technical obstacles as well \cite{Roelofs2019,Palumbo2019,Fish2020}, but a full exploration of the instrument and technological choices is beyond our goals here.

A major astrophysical uncertainty facing such an experiment is the amount of ``astrophysical noise'' present on top of the photon ring signature.  We may distinguish between two basic types of noise that might contribute to the observed visibility on the long baselines of the experiment: small-scale structures that are undoubtedly present in the turbulent flow, and long, narrow ``emission ropes'' that could in principle arise due to non-linear instabilities and temporarily mimic the photon ring.  Both kinds of noise are easy to remove if the experiment can measure absolute interferometric phase, in which case one simply averages the complex visibilities of successive measurements taken at the same points in the Fourier plane.  This reproduces the visibility of the time-averaged image, which only contains the photon ring signature, the noise having washed out.

Since measurements of absolute interferometric phase would be challenging, for practical purposes we will suppose that only the visibility amplitude is available.  The noise due to small fluctuations should still be removable by averaging the visibility amplitude (App.~\ref{app:Fluctuations}).  This averaging procedure does not reproduce the visibility amplitude of the time-averaged image, but it does preserve its periodicity, and hence the ability to infer the projected diameter $d_\varphi$ of the underlying photon ring.  Larger fluctuations caused by bright, narrow structures in the accretion flow (photon ring mimickers) would have to be modeled and excised from the data based on deviations from typical behavior in repeated measurements or other procedures.

For the initial forecast of this paper, we do not consider astrophysical fluctuations explicitly.  Rather, we assume a pristine universal signature and suppose that we have performed a 2-hour observation at each of a selection of angles $\varphi$, regarding the results as a first estimate of what might be achieved over a full mission that repeatedly observes an astrophysically noisier signal.  We find that the projected diameter $d_\varphi$ of the photon ring can be measured with a precision of $\pm0.017$~$\mu$as [Eq.~\eqref{eq:Sigma}], which provides a test of the GR prediction at the level of 0.04\% [Eq.~\eqref{eq:TestGR}].  We therefore conclude that, as shown in greater detail in the simulations of Sec.~\ref{sec:Forecast}, the answer to question 3 is: \textit{a plausible experimental configuration could make a precision test of GR.}

\section{Phenomenological models for M87*}
\label{sec:Models}

A variety of arguments indicate that M87* contains a geometrically thick, optically thin accretion disk surrounding a supermassive black hole (e.g., Ref.~\cite{EHT2019e} and references therein).  We will assume that the angular momentum of the disk is aligned with that of the black hole.  Although the flow is expected to be variable, we are interested in the time-averaged appearance and hence will consider stationary, axisymmetric emission profiles.

The M87* disk is expected to extend significantly out of the equatorial plane.  Rather than pick a specific geometric shape (or simulate an accretion flow), we will instead consider models where all the emission arises from the equatorial plane.  This can be regarded as an ``equatorial approximation'', where the polar-averaged emission from a thick disk is assigned to the relevant equatorial position.  This kind of model was used for Schwarzschild black holes with static emitters in Ref.~\cite{Gralla2019}, and here we generalize to Kerr black holes with both orbiting and infalling matter.

Although there is no systematic argument for the validity of the equatorial approximation, it has proven useful in practice.  For example, the equatorial approach predicted the presence of discrete sub-rings \cite{Gralla2019} later found in more realistic models \cite{Johnson2020}.  In this work, we similarly hope to use simple-minded models to reveal general properties, to be later confirmed in state-of-the-art modeling.

The equatorial approximation can be validated and calibrated by comparison with more realistic models.  Comparing with the results of Ref.~\cite{Johnson2020}, we find excellent qualitative agreement but notice a stark quantitative disagreement: the geometrically thick models have significantly brighter photon rings.  This may be explained by the fact that orbiting light rays spend significant time away from the equator, while still remaining in a region of high emissivity where they collect more photons.  The equatorial approximation fails to account for this effect, as it loads photons onto a light ray only at equatorial crossings.  To capture the extra brightness caused by geometrical thickness, we introduce an extra parameter that artificially enhances the flux in the photon rings.  We emphasize that this is a fudge, and as such, we call it a fudge factor.  Nevertheless, it is a highly defensible fudge, allowing the equatorial approximation to closely reproduce the time-averaged appearance of geometrically thick models: compare our Fig.~\ref{fig:PhotonRingExperiment} (left) and Fig.~\ref{fig:VisibilityAmplitude} (top right) to the analogous results presented in Fig.~1 (left) and Fig.~3 (top) of Ref.~\cite{Johnson2020}.

The advantages of the equatorial approximation are its simplicity and its speed.  It obviates the need to simulate the flow, and ray-tracing an equatorial source is vastly simpler than the full radiative transport appropriate to geometrically thick flows.  Although great strides have been made in the modeling of accretion flows and their millimeter-wave emission, we are still far from a first-principles calculation of emission, and the computational cost is steep.  The equatorial approximation bundles information from this hard work into an efficient framework in which a large variety of potential source profiles can be explored with scant computational cost.

\subsection{Description of model}

The inputs to our model are the black hole mass-to-distance ratio $M/D$, the dimensionless black hole spin $a/M$, the observer inclination $\theta_o$, the equatorial emission intensity profile $I_\mathrm{em}(r)$, and the geometrical factor $f$.  The output is a sky image and its associated complex visibility.  We now describe the model along with our computational methods.  We use units with $G=c=1$.

The first step is to ray-trace the emission profile to produce an observed image.  We use the analytic method of Refs.~\cite{GrallaLupsasca2020a,GrallaLupsasca2020b}, reviewed in App.~\ref{app:RayTracing}.  The image is described in Bardeen's ``screen coordinates'' $\alpha$ and $\beta$, which have units of black hole mass $M$.  The intensity at each screen position $(\alpha,\beta)$ is determined by tracing the associated light ray backwards through the emitting region, adding to its intensity each time it passes through the equatorial plane, and further enhancing the intensity of rays that pass more than once via the geometrical factor $f$.  Letting $r_m(\alpha,\beta)$ denote the radius at which a ray intersects the equatorial plane for the $m^\text{th}$ time on its backwards journey from screen position $(\alpha,\beta)$, we take the observed intensity to be
\begin{align}
	\label{eq:ObservedIntensity}
	I_\mathrm{obs}(\alpha,\beta)=\sum_{m=0}^{N(\alpha,\beta)}f_m\br{g(r_m,\alpha)}^4I_\mathrm{em}(r_m),
\end{align}
where $g$ is the redshift factor (observed frequency divided by emission frequency), $N+1$ is the total number of times the ray intersects the equatorial plane, and
\begin{align}
	f_m=
	\begin{cases}
		1&m=0,\\
		f&m>0.
	\end{cases}
\end{align}

We take the matter to orbit on circular geodesics outside of the Innermost Stable Circular Orbit (ISCO), and otherwise to infall on marginally stable geodesics.  The associated redshift factor $g(r,\alpha)$ was derived by Cunningham many years ago \cite{Cunningham1975}; the analytic formula is given in Eqs.~\eqref{eq:Redshift}, \eqref{eq:DiskRedshift} and \eqref{eq:GasRedshift} below.  Analytic formulas for $r_m(\alpha,\beta)$ and $N(\alpha,\beta)$ were derived recently in Ref.~\cite{GrallaLupsasca2020a}; they are given in Eqs.~\eqref{eq:EquatorialRadius} and \eqref{eq:EquatorialCrossing} below.  This analytic approach is especially convenient for photons that orbit the black hole (i.e., the photon ring photons), for which direct numerical approaches require special care.

The observed intensity $I_\mathrm{obs}(\alpha,\beta)$ computed via Eq.~\eqref{eq:ObservedIntensity} is to be understood as bolometric.  An experiment such as the EHT is instead sensitive to the specific intensity $I_\nu$ at the observation wavelength, or more directly to its Fourier transform, the complex visibility $V$.  Producing a specific intensity from our bolometric intensity requires further assumptions about the source.  We will skirt the issue and simply rescale the total intensity,
\begin{align}
	\label{eq:SpecificIntensity}
	I_\nu\propto I_\mathrm{obs}.
\end{align}
which assumes that the conversion between intensity and specific intensity is independent of image position.\footnote{An alternative approach would be to use $g^3$ in Eq.~\eqref{eq:ObservedIntensity} in place of $g^4$, which would give $I_\nu$ directly under the assumption of a broadband source.  This would produce very small differences from the method we adopt, which are degenerate with the choice of emission profile.}  We determine the proportionality coefficient by comparing with EHT observations in the visibility domain.  

The complex visibility is the Fourier transform of the specific intensity.  If the specific intensity is expressed in units of radians, with $\vec{r}=(x,y)$ denoting the image coordinate in radians, then the visibility is given by
\begin{align}
	\label{eq:ComplexVisibility}
	V\pa{\vec{u}}=\int I_\nu\pa{\vec{r}}e^{-2\pi i\vec{u}\cdot\vec{r}}\ed^2\vec{r}.
\end{align}
Here, $\vec{u}$ is the baseline vector, equal to the separation between a pair of telescopes in the plane perpendicular to the line of sight to the source, and expressed as a multiple of the observation wavelength.  We will use polar coordinates $(u,\varphi)$ for the visibility plane spanned by $\vec{u}$.

The conversion from screen coordinates $(\alpha,\beta)$ to angular distance $(x,y)$ involves the mass-to-distance ratio of the black hole, which is another input to the model.  We will scale to fiducial values via a ratio $\psi$ defined as
\begin{align}
	\label{eq:Fit}
	(M/D)_\mathrm{M87}&=\psi\cdot\pa{1.76\times10^{-11}\,\text{rad}}\\
	&=\psi\cdot\pa{3.62\,\mu\text{as}}.
\end{align}
The canonical values $M_\mathrm{M87}=6.2\times10^9M_\odot$ and $D_\mathrm{M87}=16.9$~Mpc give $\psi=1$.  If we imagine that the distance is fixed to this value, then $\psi=1$ is a mass of 6.2 billion solar masses.  This is the value favored by stellar dynamical measurements \cite{Gebhardt2011}, while the value favored by gas dynamical measurements corresponds to $\psi=0.56$ \cite{Walsh2013}.

\begin{figure}
	\centering
	\includegraphics[width=.8\columnwidth]{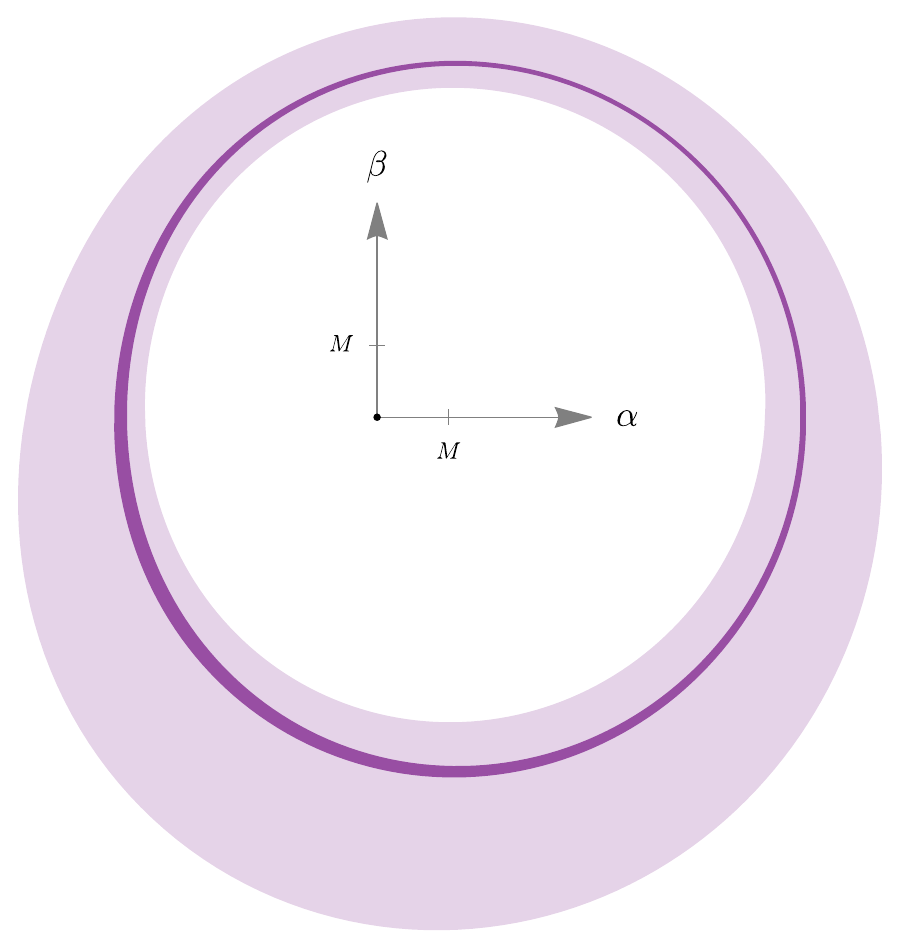}
	\caption{Illustration of the non-uniform resolution used for image computation and storage.  Defining a \textit{ray} as a complete null geodesic of the Kerr spacetime, the white, light purple, and dark purple bands correspond to rays that intersect the equatorial plane of the black hole once, twice, and three times, respectively.  The $n=1$ and $n=2$ photon rings always lie exactly within the light purple and dark purple lensing bands, respectively.  We choose the resolution of the bands such that they each contain roughly the same number of pixels.  In this example, the black hole spin is $a/M=99\%$ and the observer inclination is $\theta_o=30^\circ$.}
	\label{fig:LensingBands}
\end{figure}

\begin{figure*}
	\centering
	\includegraphics[width=\textwidth]{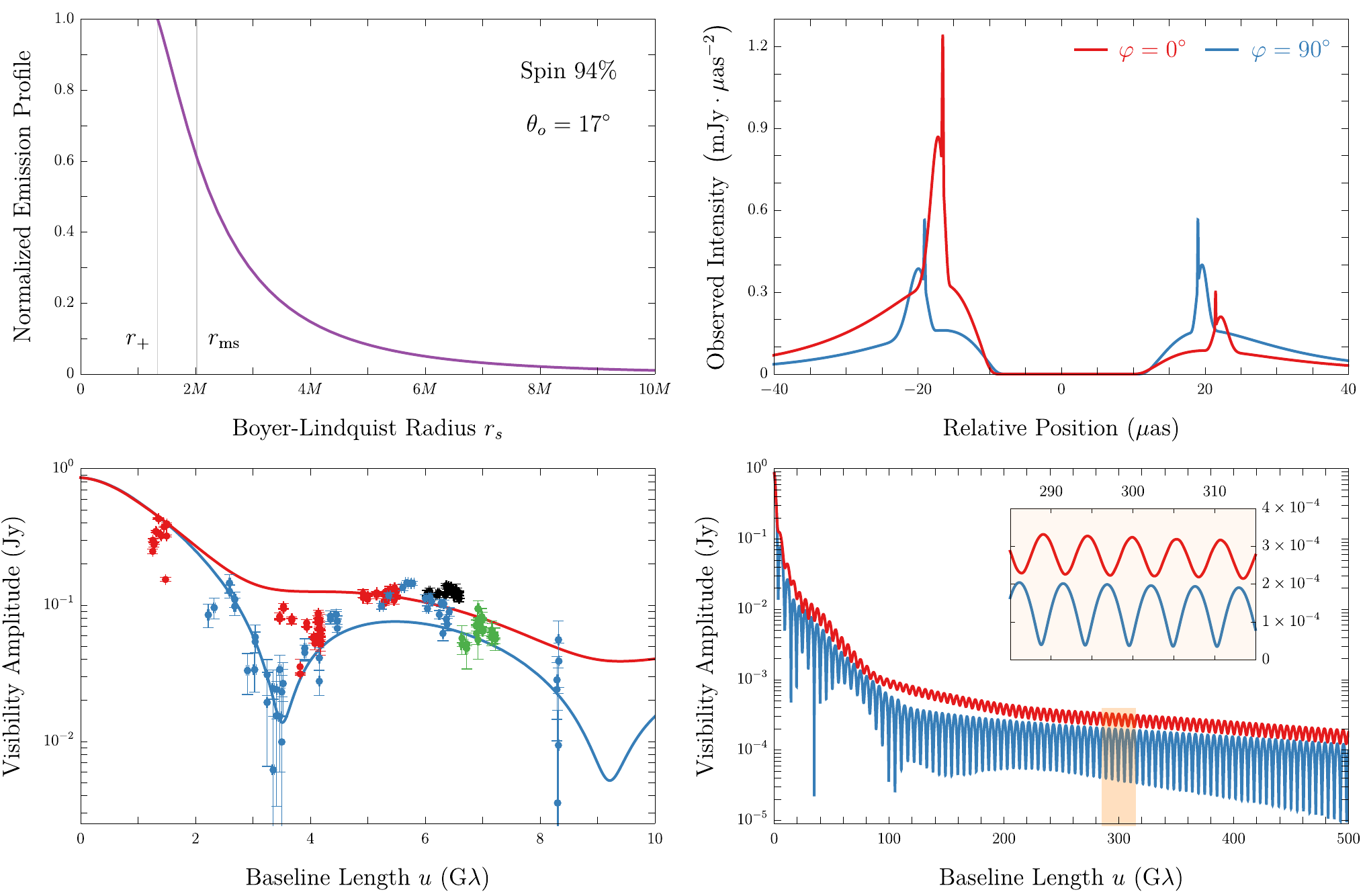}
	\caption{A best-guess model for the time-averaged appearance of M87*.  The black hole has mass consistent with stellar dynamical measurements ($\psi=1.08$) and near-maximal angular momentum ($a/M=94\%$) that is aligned with the kiloparsec-scale jet ($\theta_o=17^\circ$).  The emission originates mainly from within a few Schwarzschild radii of the event horizon (top left).  The observed intensity has a direct component, a broad $n=1$ photon ring, and a narrow $n=2$ photon ring.  (Higher order rings are not shown.)  The visibility amplitude is broadly consistent with the EHT measurements (bottom left), and shows a clean photon ring signature on long baselines (bottom right).}
	\label{fig:VisibilityAmplitude}
\end{figure*}

In principle, the above steps are straightforward: Choose model parameters, evaluate Eq.~\eqref{eq:ObservedIntensity} for the image, compute the Fourier transform \eqref{eq:ComplexVisibility} to get the visibility, and examine it on suitably long baselines.  In practice, one faces significant numerical challenges.  The essential difficulty is that the photon rings are extremely narrow compared to the overall image structure (Figs.~\ref{fig:VisibilityAmplitude} and \ref{fig:VisibilityAmplitudeBis} below).  Each ring is exponentially narrower than the last \cite{Gralla2019,Johnson2020,GrallaLupsasca2020a}, requiring exponentially fine resolution to resolve the set.  To estimate this effect, we may note that the asymptotic width ratio between successive rings ranges from about 10-20, depending on spin and inclination \cite{Johnson2020,GrallaLupsasca2020a}.\footnote{The asymptotic demagnification factor between successive rings is $e^\gamma$, where $\gamma$ is the Lyapunov exponent characterizing orbital stability in units of libration period (Mino time) \cite{Johnson2020,GrallaLupsasca2020a}.}  This suggests that the first photon ring will be 10-20 times narrower than the main structure of the image, while the second photon ring will be narrower by a factor of 100-400.  These numbers are broadly borne out by our numerical experiments.

This vast separation of scales makes it highly inefficient to work with images of uniform resolution.  Without the photon rings, a decent resolution for the main structure (primary image of the disk) might be $100\times100=10,000$ pixels.  In order to resolve the first photon ring comparably well, with a uniform resolution across the image, one would instead need millions of pixels.  Doing the same for the second photon ring (our main target) would require up to a trillion.  Instead, we adopt a non-uniform resolution adapted to the ``lensing bands'', i.e., the regions of the image for which rays orbit a given number of times before arriving at the detector (Fig.~\ref{fig:LensingBands}).  We choose roughly the same number of pixels per band, resulting in a more manageable number of pixels per image, on the order of 10 million.

\begin{figure*}
	\centering
	\includegraphics[width=\textwidth]{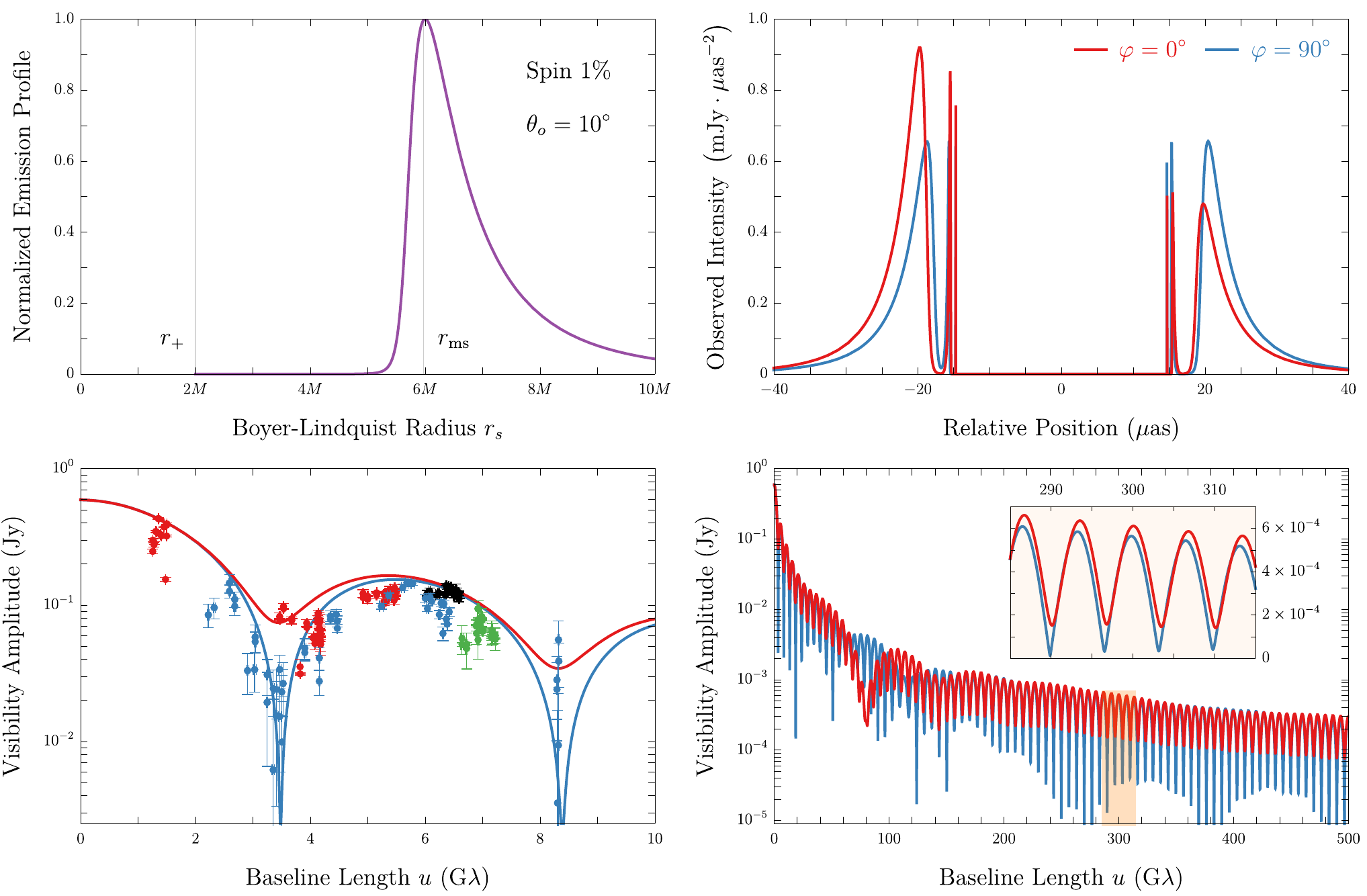}
	\caption{An alternative model designed to contrast with that of Fig.~\ref{fig:VisibilityAmplitude}.  The black hole has somewhat lower mass ($\psi=0.780$), near-zero angular momentum ($a/M=1\%$), and an accretion disk somewhat misaligned from the jet ($\theta_o=10^\circ$).  The emission originates mainly from near the ISCO at $r_\mathrm{ms}\sim6M$ (top left).  The observed intensity has a direct component, a narrow $n=1$ photon ring, and a very narrow $n=2$ photon ring.  (Higher order rings are not shown.)  The visibility amplitude is broadly consistent with the EHT measurements (bottom left), and gives a clean photon ring signature on long baselines (bottom right).}
	\label{fig:VisibilityAmplitudeBis}
\end{figure*}

Computing the Fourier transform of such an image is non-trivial.  The two-dimensional Fast Fourier Transform (FFT) requires uniform resolution, which (as we have explained) involves an intractably large number of pixels.  However, we do not need the entire Fourier transform, as we only want to check the analytic prediction \eqref{eq:UniversalSignature}, and measure $d_\varphi$ and $C_\varphi$.  For these purposes, it suffices to consider only a selection of slices through the Fourier plane at various angles $\varphi\in[0,\pi)$.  To compute the Fourier transform on these slices, we employ the projection-slice theorem.  For each angle $\varphi$, we compute the projection of the image (i.e., the integrals along lines perpendicular to the slice of constant $\varphi$ across the image), using linear interpolation across the non-uniform intensity grid.  We then Fourier transform using the one-dimensional FFT.  We check convergence by doubling the sampling resolution and ensuring that the result is unchanged.  In practice, we find that it is sufficient to sample the projection slice at intervals of $0.01M$, computing each integral along the perpendicular by sampling the intensity profile at intervals of $0.002M$.

\subsection{Functional form of emission profile}

In our model, the emission profile is a free function.  We subject ourselves only to certain general guidelines: the profile should be smooth, it should be mostly concentrated within a few Schwarzschild radii of the black hole, and it should increase as one approaches the black hole, though it may have a maximum value at some special radius (the inner edge of the disk).  We find the following functional form (derived from Johnson's SU distribution) useful for creating models of this kind:
\begin{align}
	J(r;\gamma,\mu,\sigma)=\frac{e^{-\frac{1}{2}\br{\gamma+\arcsinh\pa{\frac{r-\mu}{\sigma}}}^2}}{\sqrt{\pa{r-\mu}^2+\sigma^2}}.
\end{align}
We explore three profiles in particular:
\begin{align}
	\text{Profile 1:}\quad&
	\gamma=-\tfrac{3}{2},\
	&&\mu=r_-,
	&&\sigma=\tfrac{M}{2},\notag\\
	\text{Profile 2:}\quad&
	\gamma=0,
	&&\mu=r_-,
	&&\sigma=\tfrac{M}{2},\\
	\text{Profile 3:}\quad&
	\gamma=-2,
	&&\mu=r_\mathrm{ms}-\tfrac{M}{3},
	&&\sigma=\tfrac{M}{4}.\notag
\end{align}
As usual, $r_\mathrm{ms}$ [Eq.~\eqref{eq:ISCO}] and $r_\pm=M\pm\sqrt{M^2-a^2}$ denote the Boyer-Lindquist radii of the ISCO and outer/inner horizons, respectively.  Profiles 1 and 2 describe emission that monotonically increases as one approaches the horizon, with the latter featuring a faster rate of increase.  Profile 3 has a broad peak near the ISCO and very little emission inside.

\subsection{Two canonical models}

To illustrate the general properties of these models, we give a detailed description of two extremes: a ``best guess'' for M87* and a contrasting alternative.

Our best-guess model (Fig.~\ref{fig:VisibilityAmplitude}) is informed by the latest research on active galactic nuclei generally, and on M87* more specifically.  We suppose that the black hole spin and accretion disk are aligned with the jet, and hence adopt a canonical observer inclination of $\theta_o=17^\circ$.  Since accretion disks are expected to spin up black holes toward their maximal values \cite{Bardeen1970,Gammie2004}, we adopt a canonical value $a/M=94\%$.  (We also assume that the disk is optically thin.)  The latest phenomenological models suggest that there is significant emission near the horizon irrespective of spin \cite{EHT2019e}, and we pick our emission profile accordingly (profile 1).  Finally, we choose the geometrical factor $f=1.5$, such that the image broadly resembles the time-averaged image of Ref.~\cite{Johnson2020}.

Our alternative model (Fig.~\ref{fig:VisibilityAmplitudeBis}) deliberately ignores these research findings in order to explore the question: ``What if conventional beliefs are wrong?''  We base the contrasting model only on the simple premise that M87* should contain an accretion disk, and within this framework we purposefully stray as far as possible from the favored scenario.  We thus consider a \textit{slowly spinning} black hole with a \textit{purely equatorial} emission profile ($f=1$) terminating at the ISCO.  At present, there is no coherent theory to explain how such a disk could give rise to a relativistic jet while maintaining the relatively low luminosity of M87*, but greater surprises have occurred in the history of astrophysics.  The important point is that, even though this model strains credibility in light of current understanding, it still produces a photon ring that can be used to test general relativity.  In other words, \textit{it provides a specific example of how we could be quite wrong about the astrophysics and still test general relativity with this experiment.}  In fact, the low-spin ISCO model actually gives a much brighter signal for the proposed experiment than does the best guess model---so perhaps we should hope to be wrong!

The observational appearance of these models is shown in Figs.~\ref{fig:VisibilityAmplitude} and \ref{fig:VisibilityAmplitudeBis}.  We have used the rescaling freedom \eqref{eq:SpecificIntensity} and the choice of $M/D$ [Eq.~\eqref{eq:Fit}] to make the visibility amplitude pass roughly through the EHT observations.  The data points shown are scan-averaged visibility amplitudes from the public April 11, 2017 high-band observations, with the baseline-angle color-coding used in Fig.~1 of Ref.~\cite{EHT2019f}.  Data points colored red and blue come from baseline angles that are roughly orthogonal, making this color-coding roughly consistent with our own choice of precisely orthogonal red and blue axes.\footnote{Furthermore, the fact that our red baselines involve the highest intensity contrast is consistent with the geometric modeling and image reconstructions published by the EHT collaboration, which also have highest contrast roughly along the red direction.}  Despite their rather different image-domain appearances, both models match EHT observations quite well and produce a clear universal photon ring signature on long baselines.

\begin{figure*}
	\centering
	\includegraphics[width=\textwidth]{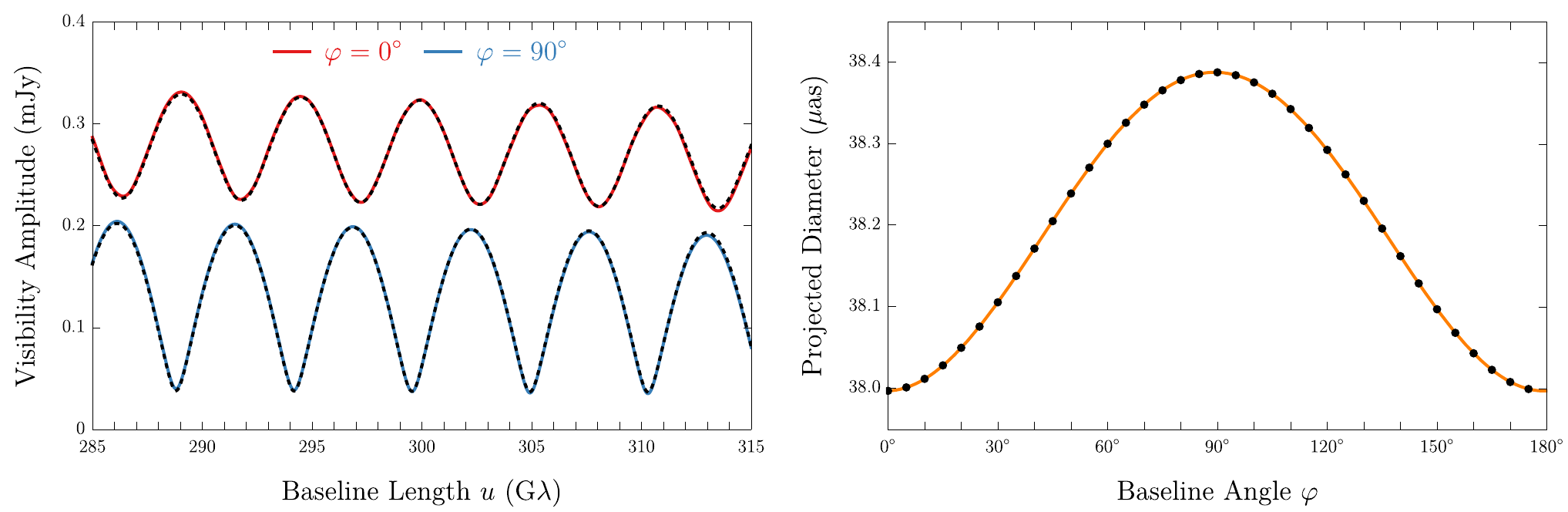}
	\caption{Determining the projected diameter $d_\varphi$ of a photon ring from its numerically computed visibility amplitude.  At each angle $\varphi$, we fit the analytic form \eqref{eq:VisibilityAmplitude} to the numerical data in the range 285-315~G$\lambda$ (left panel).  This provides the parameters $\cu{\alpha_L,\alpha_R,d}$ at every angle $\varphi$.  We then fit $d_\varphi$ to the circlipse form \eqref{eq:Circlipse} (right panel).  The circlipse fits the data extremely well, with an RMS deviation that is $4\times10^{-6}$ times the average projected diameter.  (This example is the best-guess model of Fig.~\ref{fig:VisibilityAmplitude}.)}
	\label{fig:ProjectedDiameter}
\end{figure*}

\subsection{Parameter survey}
\label{sec:Survey}

We have explored many models ($\sim$100) besides these two examples.  We now report some general findings.

First, we find that the inclination angle must be low in order to match the EHT observations.  Larger inclinations produce higher intensity contrasts, decreasing the ability of the model to reproduce the observed null at $u\sim3.5$~G$\lambda$.  Figs.~\ref{fig:VisibilityAmplitudeBis} ($\theta_o=10^\circ$) and \ref{fig:VisibilityAmplitude} ($\theta_o=17^\circ$) show the progression as the inclination is increased: the red curve becomes higher and flatter relative to the blue.  Even at $\theta_o=17^\circ$, there is larger red-blue separation than the data would suggest (Fig.~\ref{fig:VisibilityAmplitude}, bottom left), and by $\theta_o=30^\circ$, there is clear tension with the data.  We therefore deem all models with inclination larger than $30^\circ$ to be ruled out by observations.

We have conducted a systematic survey of the parameter space of viable models.  For each profile 1-3, we have considered all combinations of spins $a/M\in\cu{1\%,50\%,99\%}$, inclination angles $\theta_o\in\cu{10^\circ,20^\circ,30^\circ}$, and geometrical factors $f\in\cu{1,1.5,2}$.  We automate the choice of overall scale and $M/D$ based on EHT observations, and visually confirm the agreement in each case (although for $\theta_o=30^\circ$, it is poor).  We fit the analytic formula \eqref{eq:VisibilityAmplitude} to the numerical data on the baseline range 285-315~G$\lambda$, finding that the best-fit model deviates from the data by less than a percent.  (Performing the fit over other baseline ranges yields consistent values for $d_\varphi$---see discussion in Sec.~\ref{sec:FluxEstimate} below.)   Repeating this procedure over a range of angles $\varphi$ builds up a set of ``data points'' for the function $d_\varphi$ of interest.  We fit these data to the circlipse functional form \eqref{eq:Circlipse} and find excellent agreement: in all cases, the root-mean-square (RMS) deviation of the best-fit model is less than 0.001\%.  An example of this procedure is shown in Fig.~\ref{fig:ProjectedDiameter}, and additional results are displayed in Fig.~\ref{fig:PhotonRings} below.

The projected diameters $d_\varphi$ determined by this fitting process agree remarkably well with the $d_\varphi$ that one would infer from the image domain.  For example, the thick line illustrating $d_\varphi$ in Fig.~\ref{fig:PhotonRingExperiment} is precisely 38.138~$\mu$as in length.  This number was determined not from inspection of the image but rather from fitting the numerically computed visibility amplitude to the analytic prediction \eqref{eq:UniversalSignature}.  As can be seen by zooming in on the image, the thin ``lines of support'' extending outward pass precisely through the middle of the tiny $n=2$ photon ring at a precisely grazing angle, showing that the $d_\varphi$ measured from the universal signature \eqref{eq:UniversalSignature} is indeed the projected diameter of the second photon ring.  This demonstrates the remarkable \textit{quantitative} consistency of the various ideas underlying the proposed GR test: GR produces very thin photon rings, whose projected diameter can be measured very accurately from the long-baseline interferometric signature.

\subsection{Estimate of flux and appropriate baselines}
\label{sec:FluxEstimate}

This model parameter survey allows us to estimate the expected flux on the targeted baseline lengths of $u\sim300$~G$\lambda$.  We find that the flux is consistently in the range $\sim$0.1-1~mJy.  The only models that have flux less than 0.1~mJy feature a low-spin black hole ($a/M=1\%$) with emission very near the horizon (profiles 1 and 2) and no geometrical enhancement ($f=1$), a rather implausible combination.  Our best guess model produces flux of 0.1-0.3~mJy, shown in Fig.~\ref{fig:ProjectedDiameter}.

The largest-flux models (a few exceed 1~mJy) involve profile 3, where the emission profile has an inner edge near the ISCO.  We found this surprising at first, since these profiles generally produce much narrower photon rings as compared to their main emission (Ref.~\cite{Gralla2019} and Fig.~\ref{fig:VisibilityAmplitude}).  The explanation is that we estimate the flux at a fixed range of baselines $u\sim300$~G$\lambda$.  For models with emission near the horizon, the $n=2$ photon ring dominates at $u\sim300$~G$\lambda$, whereas for models with emission further away, it is the $n=1$ photon ring that dominates.  The $n=1$ photon ring of a low-spin ISCO model has more flux than the $n=2$ photon ring of a near-horizon model.  In general, we can expect one of these two photon rings to dominate the signal in the given range chosen for some experiment; a beating combination of the two would seem rather finely tuned.  However, observing such beats would potentially allow measurements of the relative properties of successive rings, which encode even more information about the Kerr metric \cite{Johnson2020,GrallaLupsasca2020a}.

We have targeted the baseline length $u\sim300$~G$\lambda$ as roughly the minimum telescope separation needed to ensure the presence of a universal signature.  Shorter baselines are sensitive to broader features that cannot be definitively attributed to orbiting photons.  For example, the $n=1$ photon ring in Fig.~\ref{fig:VisibilityAmplitude} has a width not too disparate from that of the main emission of Fig.~\ref{fig:VisibilityAmplitudeBis}.  For some models, it may be possible to infer a ring diameter from the periodicity on shorter baselines (e.g., Fig.~4 of Ref.~\cite{Johnson2020}), but we would be surprised if this ring diameter could be linked to properties of the underlying spacetime in a model-independent way.  By contrast, the regime $u\sim300$~G$\lambda$ that we consider is sensitive only to extremely narrow features that cannot plausibly be attributed to a persistent astronomical emission structure.  We find an excellent fit to the analytic signature \eqref{eq:UniversalSignature} near 300~G$\lambda$ for all models in our parameter survey (see discussion in Sec.~\ref{sec:Survey} above), and therefore suggest this as an appropriate target.

We have also explored fitting at 400~G$\lambda$ and 500~G$\lambda$, where we achieve similarly excellent fits with consistent values for $d_\varphi$.  The values for $\alpha_L$ and $\alpha_R$, however, depend somewhat on the fitting baseline length.  This is explained by the fact that the fall-off rate in this region is not precisely the analytically-predicted $1/\sqrt{u}$.  If an analytic fit done at (say) 300~G$\lambda$ is plotted out to (say) 500~G$\lambda$, the size of the signal does not exactly match the numerically computed results, but nonetheless the oscillation remains perfectly in phase.  This fact underlies the remarkable robustness of this method for measuring ring diameter from the universal visibility amplitude \eqref{eq:VisibilityAmplitude}.

To help quantify these claims, we provide details in one example.  The fit shown in Fig.~\eqref{fig:ProjectedDiameter} (left) was performed at 285-315~G$\lambda$.  The best-fit parameters $\cu{\alpha_L,\alpha_R,d}$ in units of $\cu{\text{Jy},\text{Jy},\mu\text{as}}$ are $\cu{149.245,27.7144,37.9974}$ for the red curve ($\varphi=0^\circ$) and $\cu{64.2793,43.8167,38.3882}$ for the blue curve ($\varphi=90^\circ$).  Performing the fit instead at 485-515~G$\lambda$, these change to $\cu{141.749,30.7711,37.996}$ and $\cu{58.6297,50.4293,38.3875}$, respectively.  That is, while $\alpha_L$ and $\alpha_R$ change by 6\%, on the other hand $d$ changes by only 0.004\%!

\subsection{Critical curve versus photon ring}
\label{sec:CriticalCurve}

\begin{figure}
	\centering
	\includegraphics[width=\columnwidth]{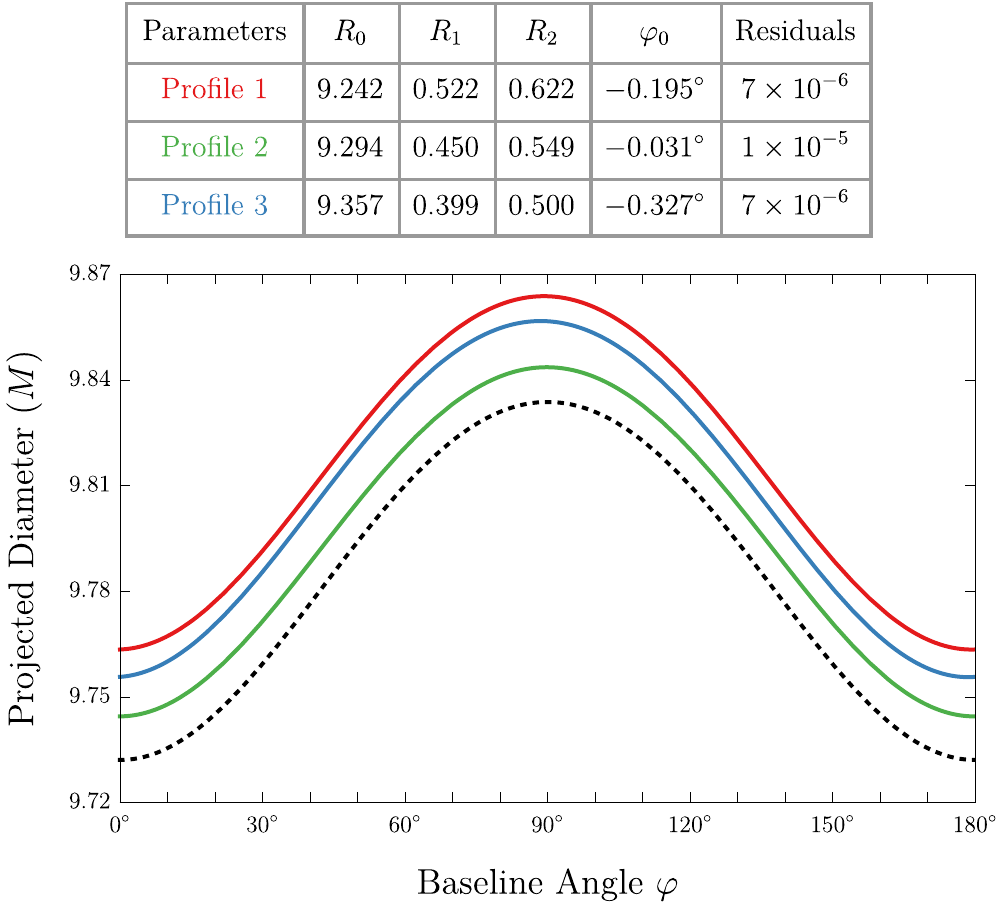}
	\caption{Comparison of observable photon rings with the critical curve of a black hole with the same parameters.  Fixing the spin  $a/M=94\%$ and inclination $\theta_o=17^\circ$, we consider the three different emission profiles 1 (red), 2 (green), and 3 (blue), all with $f=1$.  Since the black hole spins rapidly and the ISCO is near the horizon, these profiles are all quite similar, with most of the flux coming from near the horizon.  For each model, we infer $d_\varphi$ as described in Fig.~\ref{fig:ProjectedDiameter}, resulting in the parameters displayed above.  The quoted normalized residuals are RMS deviation divided by the average value of $d_\varphi$.  The curves differ in their height (astrophysics-dependent), but all share the same universal shape (GR-predicted).  The critical curve is shown in dashed black.}
	\label{fig:PhotonRings}
\end{figure}

One of the main results of this paper is the prediction that photon rings take a universal shape, whose projected diameter $d_\varphi$ we parameterize using the circlipse form \eqref{eq:Circlipse}.  We wish to emphasize that this prediction is logically distinct from shape descriptions of the critical curve, such as given recently in  Refs.~\cite{GrallaLupsasca2020c,Farah2020}.  While it has long been understood that photon rings closely follow the critical curve \cite{Falcke2000}, just \textit{how closely} has not previously been studied in quantitative detail.  Fig.~\ref{fig:PhotonRings} shows the $n=2$ photon ring diameter for three different emission profiles around the same black hole, along with the critical curve for that black hole.  While the precise curves differ, their shapes are clearly similar, and this is quantified to a part in $10^5$ by the fact that all four are described by the circlipse family of shapes \eqref{eq:Circlipse}.

Although studying the critical curve led us to to the circlipse family \cite{GrallaLupsasca2020c}, the shape result for the photon rings is logically distinct.  One cannot directly observe the critical curve of a black hole, and the black hole parameters cannot (in practice) be known well enough to infer the critical curve.  In this sense, comparing photon ring and critical curve in a single model is of purely theoretical interest.  A logical leap taken in this paper is that, for testing GR, we do not even \textit{care} about the critical curve of the black hole we observe; we only need to check the shape of its photon rings.

\begin{figure*}
	\begin{minipage}[b]{1.02\columnwidth}
		\includegraphics[width=\textwidth]{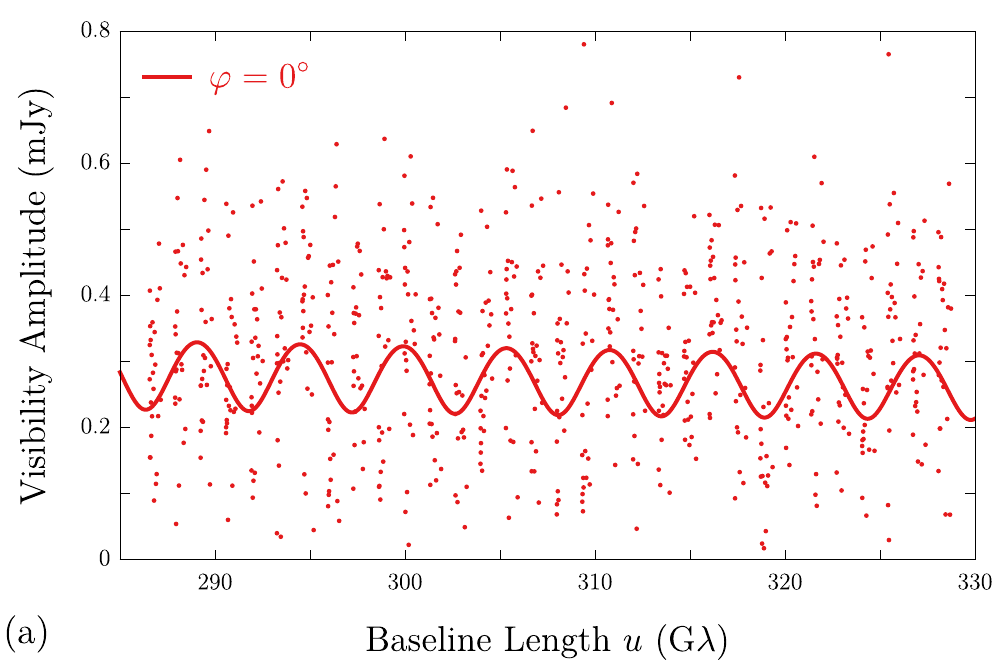}
	\end{minipage}
	\hfill
	\begin{minipage}[b]{1.02\columnwidth}
		\includegraphics[width=\textwidth]{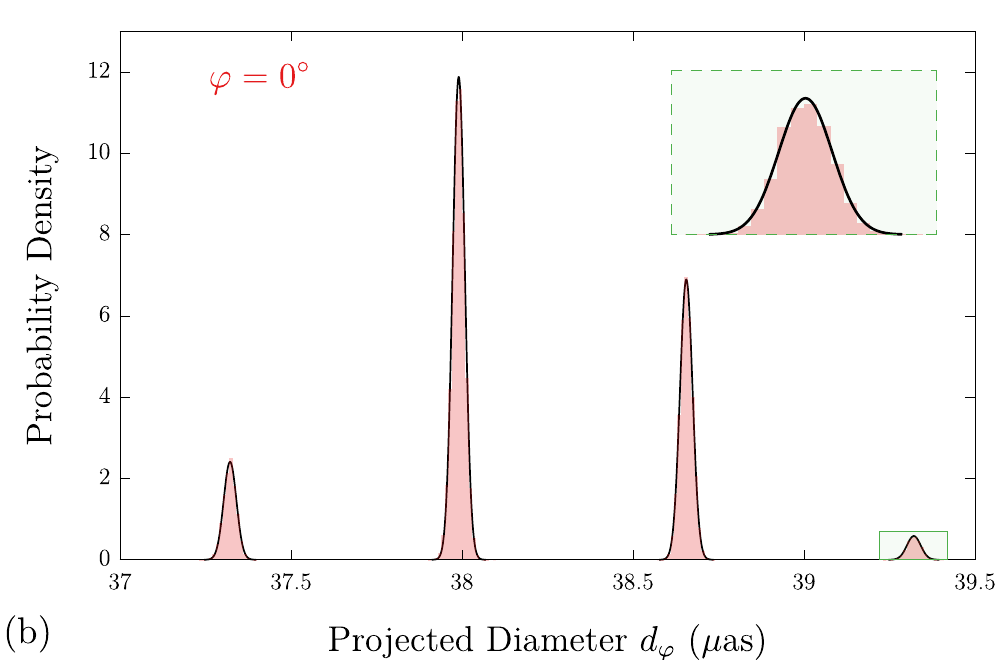}
	\end{minipage}
	\begin{minipage}[b]{1.02\columnwidth}
		\includegraphics[width=\textwidth]{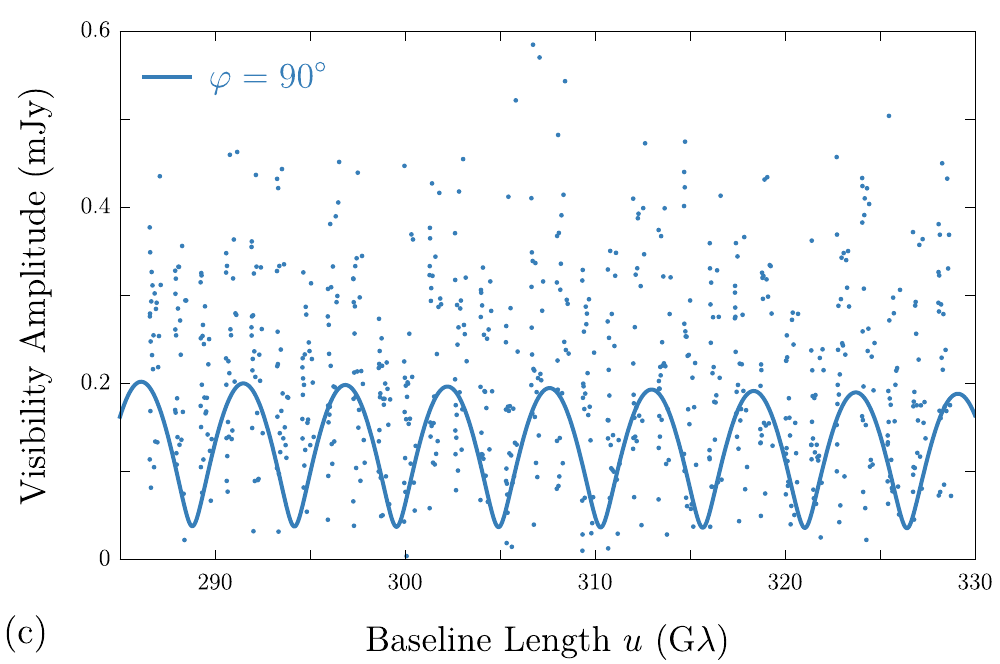}
	\end{minipage}
	\hfill
	\begin{minipage}[b]{1.02\columnwidth}
		\includegraphics[width=\textwidth]{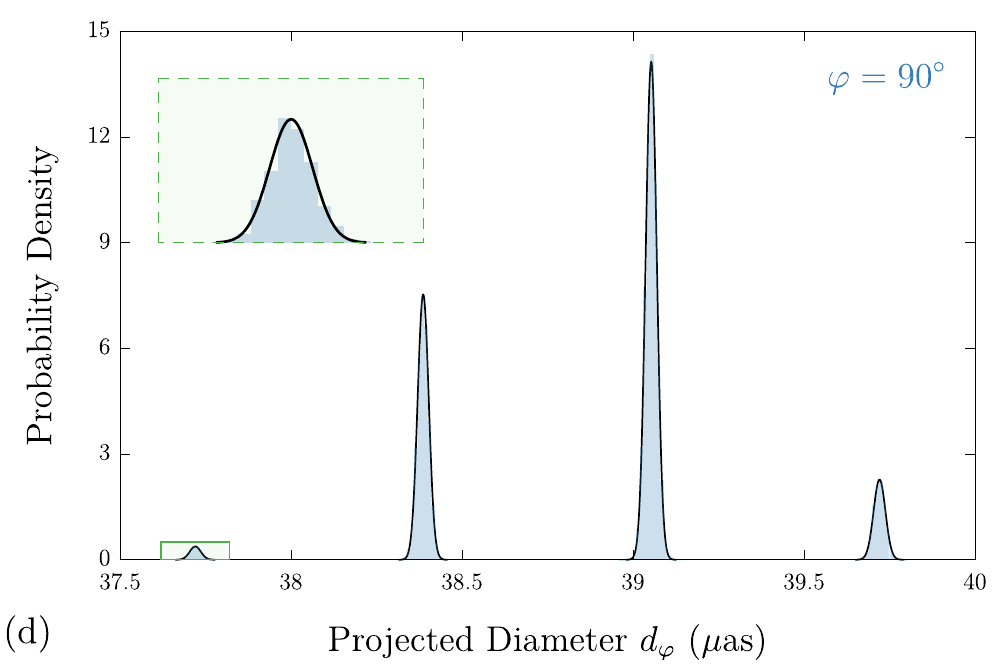}
	\end{minipage}
	\begin{minipage}[b]{1.02\columnwidth}
		\includegraphics[width=\textwidth]{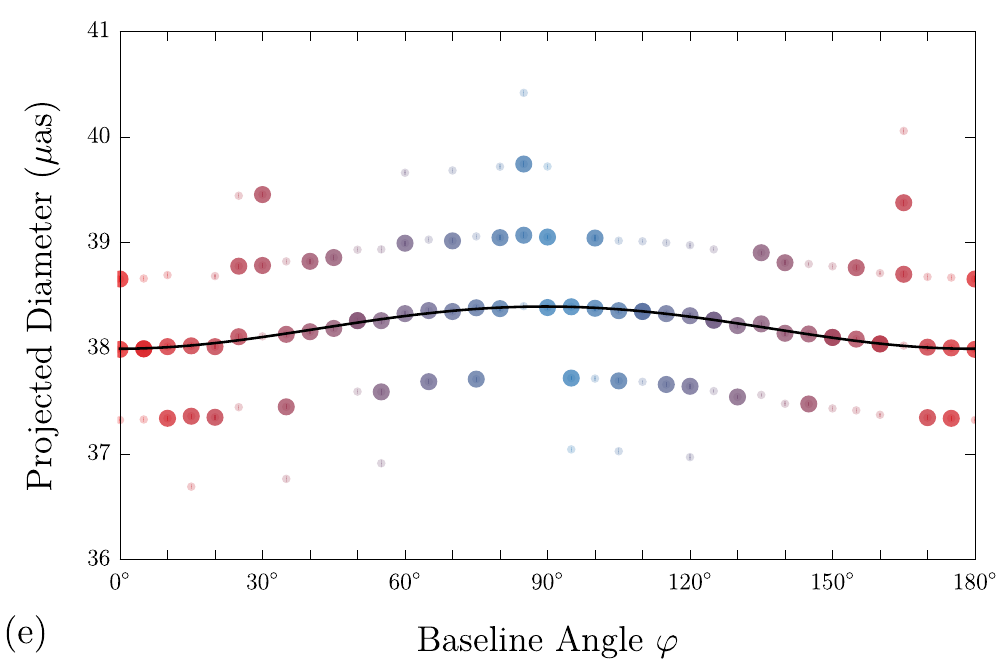}
	\end{minipage}
	\hfill
	\begin{minipage}[b]{\columnwidth}
		\caption{Illustration of the data analysis method used in the experimental forecast.  Mock data is generated at each angle $\varphi$, producing spiky probability distributions on $d_\varphi$ [panels (a)-(d)].  We fit each spike to a Gaussian and determine its central value, standard deviation, and total probability.  This information is shown in panel (e).  We have included only points that together contain at least $95\%$ of the probability at each angle $\varphi$, and the subset that contain at least $68\%$ of the probability are overplotted with larger dots.  We determine the best-fit circlipse model \eqref{eq:Circlipse} (black line) by maximizing the likelihood, which is the product of the 36 multi-peaked likelihoods at each angle.  A zoom-in of the best-fit circlipse is shown in the right panel of Fig.~\ref{fig:PhotonRingExperiment}.\vspace{22pt}}
		\label{fig:MCMC}
	\end{minipage}
\end{figure*}

\section{Forecast}
\label{sec:Forecast}

The analysis of the previous section identifies the range $u\sim300$~G$\lambda$ as a promising target for photon ring shape tests, with the expected flux ranging from 0.1~mJy to 1~mJy.  We now consider a simple mock experiment in order to forecast what precision might be achieved in practice.  In this paper, we confine ourselves to the analysis of a single canonical configuration, aiming to establish ballpark experimental targets for sensitivity and bandwidth, and also to illustrate the type of analysis needed to transform interferometric measurements into tests of general relativity.

Our canonical configuration was described in Sec.~\ref{sec:Measurable}.  It involves an orbiting space telescope (with an orbital period of 720 hours, or $\sim$1 month) as well as a ground station, and measures 768 complex visibilites in the range $u\sim285$-330~G$\lambda$ over each 2-hour ``observing run''.  Each measured visibility is assumed to have an experimental error of 0.14~mJy.  We suppose that the source has been observed at $\varphi=\cu{0^\circ,5^\circ,\dots,175^\circ}$, corresponding to 36 two-hour observations taken over the course of approximately one month.\footnote{Throughout this paper, we have worked with angles separated by $5^\circ$ in order to obtain round numbers not tied to the details of any particular experimental forecast.  For the forecast of this section, a more convenient choice would be to collect data every $12^\circ$ around the orbit, so that observations are separated by 24 hours, and the 15 data points collected over two weeks.  (Since $\varphi$ and $\varphi+180^\circ$ represent the same physical $d_\varphi$, all projected diameters are measurable over any half-orbit of the satellite.)  If 36 data points separated by $5^\circ$ are instead desired, a realistic observing strategy would take slightly over a month.}  We create mock data assuming that the true underlying signal is that of our best-guess model (Fig.~\ref{fig:VisibilityAmplitude}).  The resulting forecast yields the precision that can be achieved in about a month in the best-case scenario that the signal is pure, with no need to average away astrophysical fluctuations.  In the more likely scenario that repeated measurements at the same $\varphi$ are necessary to reveal the photon ring signature, our forecast still provides a basic estimate of what could be achieved over a longer-duration mission.

Our approach is illustrated in Fig.~\ref{fig:MCMC}.  At each angle $\varphi\in\cu{0^\circ,5^\circ,\dots,175^\circ}$, we generate mock data (Fig.~\ref{fig:MCMC}\textcolor{red}{(a),(c)}) by adding independent complex Gaussian noise ($\sigma=0.14$~mJy) to the numerically computed complex visibility at each of the 192 visibility points measured by the mock experiment.  We sample the likelihood function (using a Rice distribution with $\sigma=0.14$~mJy to describe the visibility amplitudes) using a Markov Chain Monte Carlo (MCMC) approach.  We impose no Bayesian priors and therefore regard the likelihood as the probability density function on the model parameters $\cu{\alpha_L,\alpha_R,d}$.  We find that $\alpha_L$ and $\alpha_R$ are in general poorly constrained.  However, marginalizing over these parameters reveals a sharp, multi-peaked probability distribution for $d$ at every angle $\varphi$ (Fig.~\ref{fig:MCMC}\textcolor{red}{(b),(d)}).  

The multi-peaked distribution visible in Fig.~\ref{fig:MCMC}\textcolor{red}{(b),(d)} arises because the signature \eqref{eq:VisibilityAmplitude} of the photon ring does \textit{not} contain a parameter for the phase of the oscillation---only its period $1/d_\varphi$ and maximum and minimum values (simply related to $\alpha_L$ and $\alpha_R$) appear in the formula.  Determining the projected diameter $d_\varphi$ from the analytic fit effectively counts the number of hops between $u=0$ and $u\sim300$~G$\lambda$.  The different allowed values of $d_\varphi$ correspond to different numbers of hops, such that the phase of the oscillation remains the same at $u\sim300$~G$\lambda$.  Since Eq.~\eqref{eq:VisibilityAmplitude} depends only on the combination $d_\varphi u$ (over a sufficiently small range of baselines), the spacing between acceptable values of $d_\varphi$ is approximately $1/u$.  For our range $u\sim300$~G$\lambda$, we have $1/u\approx0.68$~$\mu$as, which perfectly matches the measured spacing between the peaks.

Our MCMC sampling provides a multi-peaked distribution on $d_\varphi$ for each angle $\varphi$ in the set we consider.  For each $\varphi$, we individually fit each peak of the distribution to a Gaussian, providing the central value, width, and total probability associated with each peak.  We display the central values and widths ($1\sigma$ error bars) in Fig.~\ref{fig:MCMC}\textcolor{red}{(e)}.  Each point on this plot also comes with a total probability $P$, equal to the area under the peak that it represents.  At each angle $\varphi$, we have included the bumps that together contain at least $95\%$ of the probability, and we have overplotted large circles on the subset that together contain at least 68\% of the probability.\footnote{Note that because of the multi-peaked nature of the probability distribution, the set of large circles at each $\varphi$ often contains significantly more than $68\%$ of the total probability.}  In Fig.~\ref{fig:MCMC}\textcolor{red}{(e)}, each vertical line of points can be understood as a top-down view of the spiky probability distribution at the relevant angle $\varphi$, shown from the side for $\varphi\in\cu{0^\circ,90^\circ}$ in Fig.~\ref{fig:MCMC}\textcolor{red}{(b),(d)}.

\subsection{GR test}

To compare the results of our mock experiment against the GR prediction, we use the maximum likelihood method to determine the circlipse \eqref{eq:Circlipse} that best fits the data represented in  Fig.~\ref{fig:MCMC}\textcolor{red}{(e)}.  Each individual likelihood (at a given angle $\varphi$) is expressed as the sum of the Gaussians determined in the fitting process described above.  The total likelihood of the data is then the product of these 36, individually multi-peaked likelihood functions.  The maximum likelihood circlipse is shown in black.

By eye, it may appear that other acceptable fits could be found simply by shifting the black curve up or down, so that it passes through a different ``horizontal band'' of points.  However, the likelihood scales with the product of the probabilities of the points that the curve passes through, which one can easily see will be many orders of magnitude smaller.  Indeed, we find local maxima of the likelihood associated with curves that are shifted up or down, and these are all at least fifteen orders of magnitude less likely than the best fit.  In practice, this means that for the purpose of fitting a model to the data, it suffices to consider only the middle band of data points, as it is (effectively) certain to be the correct one.  We therefore show only these points in  Fig.~\ref{fig:PhotonRingExperiment} as the main result of the mock experiment.

To characterize the precision of the measurement, we report the average standard deviation of this middle band of points:
\begin{align}
	\label{eq:Sigma}
	\sigma=0.017\ \mu\textrm{as}.
\end{align}
We could just as well compute a probability-weighted average of the standard deviations of all the data points in Fig.~\ref{fig:MCMC}\textcolor{red}{(e)}, or indeed an unweighted average: this figure is robust to reasonable changes in its definition.

To assess the goodness of fit, we discard the probability information and consider all points in the preferred band to be equally likely.  This enables us to use the standard reduced chi-squared metric,
\begin{align}
	\chi^2_r=\sum_{i=1}^N\frac{1}{N-m}\pa{\frac{d_{\mathrm{obs},i}-d_{\mathrm{GR},i}}{\sigma_i}}^2,
\end{align}
where $N=36$ is the number of data points (angles $\varphi$), and $m=4$ is the number of model parameters.  For the best-fit circlipse, we obtain $\chi^2_r=0.95$.  

We also fit to the simpler ellipse model, which is equivalent to the circlipse with $R_0=0$ and thus has $m=3$ parameters.  The resulting fit yields $\chi^2_r=0.97$.  That is, for our best-guess model and this experimental forecast, the measurement precision is inadequate to distinguish between the ellipse description of the true curve and the more accurate circlipse description.  The ellipse model is more convenient for reporting parameters, since its parameters are not degenerate.  We find the parameters $R_1=38.001\pm.005$~$\mu$as, $R_2=38.400\pm.005$~$\mu$as and $o=-0.6\pm0.5^\circ$.  The parameters $R_1$ and $R_2$ represent the two radii of the ellipse, while the angle $o$ indicates that it is slightly rotated relative to the Bardeen coordinates for the image.  The shorter axis is the (nearly) horizontal one, perpendicular to the projected spin axis.

Based on these chi-square values, we determine that the GR model is an adequate fit to the data, i.e., GR ``passes the test''.  We report the precision of the test using the normalized RMS residuals of the best-fitting circlipse,
\begin{align}
	\label{eq:TestGR}
	\frac{\sqrt{\left\langle\pa{d_\mathrm{obs}-d_\mathrm{GR}}^2\right\rangle}}{\langle d_\mathrm{GR}\rangle}=0.04\%.
\end{align}
This number is unchanged if we instead use the simpler, ellipse model for the fit.  If an experiment actually measured these mock data points, we would say that it confirmed the GR shape prediction at the 0.04\% level.  If instead the measurement produced data points for which the fit is inadequate (measured with a large reduced chi-squared), we would report a deviation from general relativity at the associated $p$-value.

We have assumed in this forecast that 36 angles $\varphi$ were measured, but GR tests are still possible with far fewer data points.  In principle, one needs only five measurements, since the model family \eqref{eq:Circlipse} has four parameters.  With so few points, and depending on the final level of experimental and astrophysical noise, it is possible that there will be two comparably likely circlipse fits, separated by a shift up or down to another ``horizontal band'' in Fig.~\ref{fig:MCMC}\textcolor{red}{(e)}.  However, one can still test the GR shape prediction without knowing which curve is the true one, simply by confirming that both fits are adequate.  For the present forecast, we may see from  Fig.~\ref{fig:MCMC}\textcolor{red}{(e)} that measuring 5 points is typically enough to strongly prefer one band, and that adding even a few more would make the preference overwhelming.

\subsection{Mass measurements}

Although we have focused on testing GR, the data from the proposed experiment can also be used to measure black hole parameters (assuming the correctness of GR).  In particular, one would like to be able to measure the black hole mass $M$, its spin parameter $a$, and the observer inclination $\theta_o$.  Forecasting the accuracy of such a measurement requires further theoretical work exploring the correlation between the photon ring projected diameter $d_\varphi$ and the underlying black hole parameters. 

However, we can gain a rough estimate of the expected precision by working under the approximation that the $n=2$ photon ring agrees with the critical curve, which holds in our models at around the percent level (Fig.~\ref{fig:PhotonRings}).  The size of the critical curve is set primarily by the black hole mass, with additional 5-10\% variation depending on spin and inclination (e.g., Fig.~7 of Ref.~\cite{Johnson2020}).  One can therefore measure the mass within a 5-10\% uncertainty, with improvements in precision if the spin and inclination can be independently determined.  However, spin and inclination affect the shape of the critical curve by at most $\sim 1\%$ at the modest inclinations presumed for M87*, which is the same order of magnitude at which differences between the $n=2$ photon ring and the critical curve become important.  Further work is required to estimate the precision of potential measurements of spin and inclination and to improve the mass forecast beyond the conservative estimate of 5-10\% accuracy.

\section{Comparison with other work}
\label{sec:Comparison}

There has been some previous discussion of testing GR with interferometric observations on Earth-sized baselines \cite{Psaltis2019}.  The proposed precision of such a test is $\sim$10\% for observations of Sgr~A*  \cite{Psaltis2015}. This estimate is subject to significant astrophysical uncertainty, because the method relies on the identification of a ring-like feature with the critical curve, without having significant ability to select for the sharp lensing features that we focus on here.  To regard such a measurement as evidence against GR, one has to have a prior belief in the correctness of the astrophysical assumptions that is at least comparable to one's prior belief in the correctness of GR, one that will be hard to support from the data themselves.

A more general way of phrasing the difficulty with ground-based GR tests is that the majority of the photons contributing on these baselines have not orbited the black hole \cite{Gralla2019,Johnson2020}.  These photon trajectories bend only modestly in the vicinity of the black hole \cite{GrallaLupsasca2020a} and are not sensitive to any of its detailed properties.\footnote{For example, for direct emission reaching a nearly face-on observer, the arrival radius $b=\sqrt{\alpha^2+\beta^2}$ of a photon emitted from equatorial radius $r_s$ can be calculated to good accuracy, regardless of the black hole spin, by ``just adding one'': $b\approx r_s+M$  \cite{GrallaLupsasca2020a}.}   

The photon ring test discussed herein does not suffer from these drawbacks.  The visibility signature is unambiguous and cannot be mimicked by any astrophysically plausible source.  If the signature is detected, it will provide an unambiguous and quantitatively precise test of the Kerr black hole prediction.

In this paper, we have considered a test of GR on its own terms, without a comparison theory.  Of course, data from the proposed photon ring experiment can also be used for model comparison.  The alternative model could derive from a modified theory of gravity (e.g., Refs.~\cite{Cunha2017,Okounkova2019}) or from a parameterized deviation away from the Kerr metric (e.g., Refs.~\cite{Johannsen2010,Vigeland2011,Medeiros2020,Carson2020} and references therein).

This test of strong-field GR will be complementary to ongoing black-hole merger tests \cite{LIGO2016b}, which will surely become more precise as detector sensitivity improves and further loud sources are detected.  Merger tests probe the violent, highly dynamical spacetime of black holes colliding, whereas the photon ring reflects the stable, pristine gravity of the Kerr black hole.  The merger tests probe the generation of gravitational disturbances by the motion of heavy masses, while the photon ring test probes the extreme bending of massless electromagnetic waves.  Together, these tests cover opposite extremes of strong-field phenomena.

\section{Outlook}
\label{sec:Summary}

We have argued that M87* is a promising target for a new and  exceptionally precise test of strong-field general relativity.  The main limitations of our work arise from the class of simplified, equatorial models that we consider.  Although we stand by the genericity of our predictions, we also recognize the importance of reexamining them in the context of more realistic models involving fluctuating, geometrically thick structures.

Perhaps the most important limitation of our analysis is that it gives little indication of the size and scale of contaminating ``astrophysical noise'' that must be averaged away.  Although the analysis undertaken so far indicates that the mission described here can indeed detect the universal signature \eqref{eq:UniversalSignature} and thereby test GR to high precision, it is difficult to estimate the exact precision as a function of experimental parameters/mission design without better constraints on the astrophysical noise level. 

Along those lines, we would like to emphasize that our fiducial experimental configuration is merely a convenient choice for a first demonstration that this kind of measurement may indeed be possible in practice.  In this configuration, we have demanded a rather low thermal noise level that remains beyond the reach of currently demonstrated space antenna and receiver technologies at millimeter wavelengths.  Though the sensitivity requirement is almost certainly critical to this measurement, many additions to the experimental design could be envisioned to improve the measurement, make the fringe-finding more viable, increase the instantaneous or short-timescale baseline coverage, or otherwise make the experiment more robust. A detailed concept is deferred to future work.

At the end of a theoretical paper proposing a novel experiment, it is appropriate to ask: If the experiment were built to specification, what could go wrong?  One possibility is that there simply is not enough flux at 300~G$\lambda$.  We are confident in our prediction of at least a few tenths of a mJy (see also Ref.~\cite{Johnson2020}), but we also expect that the accuracy of the prediction will improve dramatically with focused efforts on the modeling side as well as additional horizon-scale observations.  Another possibility is that there may be more astrophysical noise than expected. We hope this could be mitigated with flexible mission design, allowing for additional observations beyond those originally planned, in order to average down the noise.  A third possibility is that the photon ring is suppressed by synchrotron opacity in the inner accretion flow, in which case no measurement is possible.  While more work is needed, all published state-of-the-art simulations of M87* to date reveal a prominent photon ring feature \cite{EHT2019e}.

Notice, however, that none of these various failure modes involves an astrophysical effect mimicking the photon ring signature, or otherwise allows one to be fooled into thinking that GR has been tested when in reality it has not.  This is why we find the experiment so compelling: if it works, we have a true confrontation between theory and experiment.  In other words, \textit{it is possible to be surprised}.  Not merely an exercise in confirmation of expectations, the proposed experiment offers a clear framework for finding a deviation from GR.

\section*{Acknowledgements}

SG thanks Megan Gralla for helpful conversations.  AL gratefully acknowledges support from the Jacob Goldfield Foundation.  This work was supported in part by NSF grant PHY-1752809 to the University of Arizona.

\appendix

\section{Averaging over fluctuations}
\label{app:Fluctuations}

\begin{figure*}
	\centering
	\includegraphics[width=\textwidth]{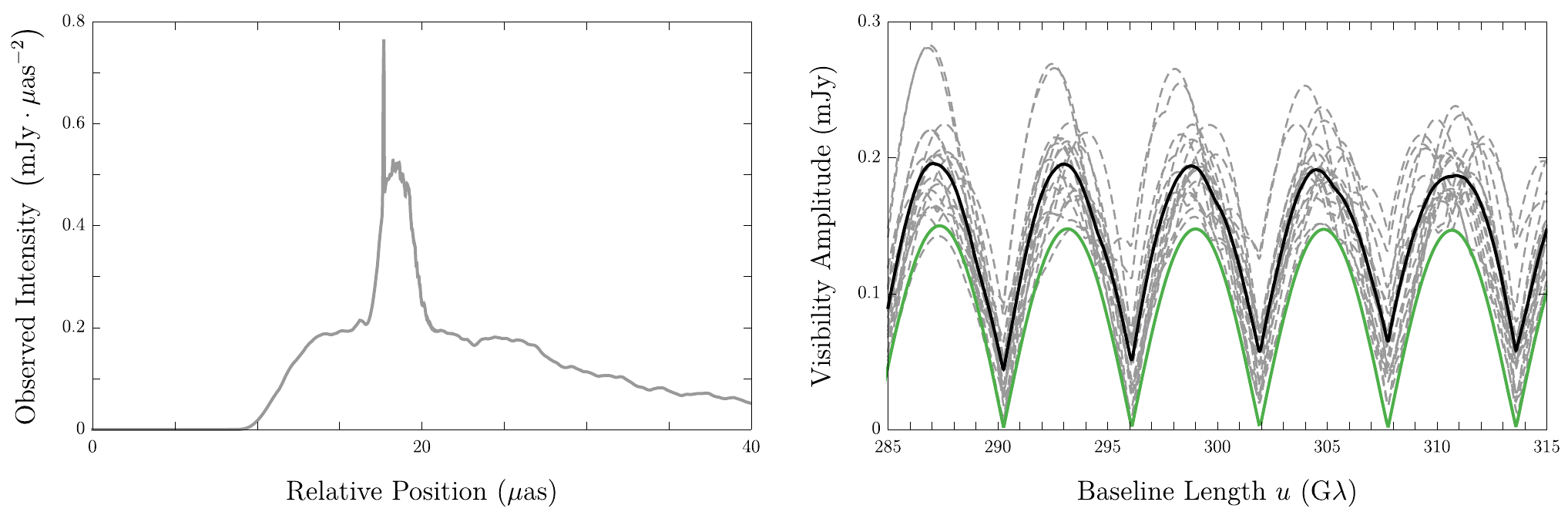}
	\caption{Averaging over fluctuations in the visibility amplitude.  On the left, we show the observed intensity of a stationary, axisymmetric model that has been augmented with noise.  On the right we show the long-baseline visibility amplitude of this model, together with seventeen other realizations of the noise added to the same underlying model (dashed gray).  The averaged visibility amplitude (black) clearly shows the same periodicity of the underlying noise-free model (green).}
	\label{fig:NoisyCase}
\end{figure*}

An important limitation of this work is our use of stationary, axisymmetric source profiles, which are unable to model the expected variability in the observed visibility.  Although future work is necessary to fully address this shortcoming, in this appendix we illustrate how variability might be dealt with in practice.  We consider for simplicity the case of a face-on observer, making the observed image axisymmetric.  We simulate source fluctuations by adding random (axisymmetric) noise to otherwise smooth intensity profiles $I_\mathrm{em}(r)$.  We adjust the noise properties so that the resulting image cross-sections are visually similar to published cross-sections arising from GRMHD simulations (right panel in Fig.~4 of Ref.~\cite{Psaltis2015}).  We settle on the formula
\begin{align}
	I_\mathrm{em}(r)=I_0(r)\br{1+0.05\,\mathcal{N}(r)},
\end{align}
where $I_0(r)$ is some smooth intensity profile and $\mathcal{N}(r)$ is constructed by smoothing (via a  moving average over width $0.1M$) a fractional Gaussian noise process in increments of $0.01M$, choosing zero mean, covariance 1.5, and Hurst index 0.5.  An example is shown in Fig.~\ref{fig:NoisyCase} left.

Although this noise level is chosen for the sake of consistency with cross-sections of GRMHD-simulated images, we emphasize that our noise is fully axisymmetric, such that each fluctuation corresponds to a whole \textit{ring} of noise.  This is a highly adversarial kind of noise, since such ``noise rings'' most directly threaten to contaminate the photon ring signal.  We therefore view our noise estimate for the visibility amplitude as conservative, and most likely an overestimate of the true noise.  Nevertheless, Fig.~\ref{fig:NoisyCase} shows that even in this case, by averaging the visibility amplitude over many successive observations, one obtains a clear periodic signal whose nulls agree precisely with the underlying noise-free model.  We conclude that photon ring diameter measurements are not significantly impacted by this level of astrophysical noise.

\section{Redshift factor}
\label{app:Redshift}

Our equatorial approximation assigns an emission intensity to every equatorial radius in the Kerr spacetime.  To compute the observed intensity, we must take into account the relative redshift between source and detector.  We will use the redshift factor of circular orbiters down to the ISCO, beyond which we will take the matter to infall along geodesics with conserved quantities equal to those of the last stable orbit.  The redshift factor for such a source was worked out by Cunningham in 1975 \cite{Cunningham1975}.  We review the derivation and present the results.

The redshift factor $g$ is defined as the observed frequency over the emitted frequency.  For the  equatorial orbits of interest, it depends only on the source radius $r_s$ and the conserved quantity $\alpha$.  We express the results as
\begin{align}
	\label{eq:Redshift}
	g(r_s,\alpha)=
	\begin{cases}
		g_\mathrm{orbit}&r\ge r_\mathrm{ms},\\ 
		g_\mathrm{infall}&r<r_\mathrm{ms},
	\end{cases}
\end{align}
where $r_\mathrm{ms}$, $g_\mathrm{orbit}$ and $g_\mathrm{infall}$ are respectively given (in terms of $\lambda=-\alpha\sin{\theta_o}$) in Eqs.~\eqref{eq:ISCO}, \eqref{eq:DiskRedshift} and \eqref{eq:GasRedshift} below.

A particle orbiting a Kerr black hole on a prograde, circular, equatorial geodesic at Boyer-Lindquist radius $r=r_s$ has four-velocity \cite{Bardeen1972}
\begin{align}
	\label{eq:EquatorialGeodesics}
	u_s&=u_s^t\pa{\pd_t+\frac{M^{1/2}}{r_s^{3/2}+aM^{1/2}}\pd_\phi},\\
	u_s^t&=\frac{r_s^{3/2}+aM^{1/2}}{\sqrt{r_s^3-3Mr_s^2+2aM^{1/2}r_s^{3/2}}}.
\end{align}
Such a geodesic has energy and angular momentum
\begin{align}
	\label{eq:ConservedEnergy}
	\frac{E}{\mu}&=\frac{r_s^{3/2}-2M\sqrt{r_s}+a\sqrt{M}}{r_s^{3/4}\sqrt{r_s^{3/2}-3M\sqrt{r_s}+2a\sqrt{M}}},\\
	\label{eq:ConservedAngularMomentum}
	\frac{L}{\mu}&=\frac{\sqrt{M}\pa{r_s^2-2a\sqrt{Mr_s}+a^2}}{r_s^{3/4}\sqrt{r_s^{3/2}-3M\sqrt{r_s}+2a\sqrt{M}}},
\end{align}
and remains stable provided that $r_s\ge r_\mathrm{ms}$, where the ISCO radius $r_\mathrm{ms}$ is given by
\begin{subequations}
\label{eq:ISCO}
\begin{align}
	r_\mathrm{ms}&=M\pa{3+Z_2-\sqrt{\pa{3-Z_1}\pa{3+Z_1+2Z_2}}},\\
	Z_1&=1+\sqrt[3]{1-a_\star^2}\br{\sqrt[3]{1+a_\star}+\sqrt[3]{1-a_\star}},\\
	Z_2&=\pa{3a_\star^2+Z_1^2}^{1/2},\quad
	a_\star=\frac{a}{M}.
\end{align}
\end{subequations}
The ratio $g=\nu_\mathrm{obs}/\nu_\mathrm{em}$ may be computed
using Kerr photon conserved quantities (e.g., Refs.~\cite{Cunningham1973,Gralla2018}) to be 
\begin{align}
	\label{eq:DiskRedshift}
	g_\mathrm{orbit}=\frac{\sqrt{r_s^3-3Mr_s^2+2a\sqrt{M}r_s^{3/2}}}{r_s^{3/2}+\sqrt{M}\pa{a-\lambda}},
\end{align}
Within the ISCO, we assume that gas flows along equatorial geodesics with the same conserved quantities as the ISCO, i.e., Eqs.~\eqref{eq:ConservedEnergy} and \eqref{eq:ConservedAngularMomentum} evaluated at $r_s=r_\mathrm{ms}$.  Together with the vanishing of the Carter constant (required of all equatorial timelike geodesics), these conditions may be used to algebraically derive the four-velocity and compute the redshift \cite{Cunningham1975}.  In the notation thereof, the results are
\begin{align}
	u&=u^r\pd_r+u^\phi\pd_\phi+u^t\pd_t,\\
	u^r&=-\sqrt{\frac{2}{3}\frac{M}{r_\mathrm{ms}}}\pa{\frac{r_\mathrm{ms}}{r_s}-1}^{3/2},\\
	u^\phi&=\frac{\gamma_\mathrm{ms}}{r_s^2}\pa{\lambda_\mathrm{ms}+aH},\\
	u^t&=\gamma_\mathrm{ms}\br{1+\frac{2M}{r_s}\pa{1+H}},\\
	H&=\frac{2Mr_s-a\lambda_\mathrm{ms}}{\Delta(r_s)},
\end{align}
where
\begin{align}
	\lambda_\mathrm{ms}&=\frac{\sqrt{M}\pa{r_\mathrm{ms}^2-2a\sqrt{Mr_\mathrm{ms}}+a^2}}{r_\mathrm{ms}^{3/2}-2M\sqrt{r_\mathrm{ms}}+a\sqrt{M}},\\
	\gamma_\mathrm{ms}&=\sqrt{1-\frac{2}{3}\frac{M}{r_\mathrm{ms}}}.
\end{align}
The final result for the redshift factor is
\begin{align}
	\label{eq:GasRedshift}
	g_\mathrm{infall}=\frac{1}{u^t-u^\phi\lambda-u^r\br{\Delta(r_s)}\br{\pm\sqrt{\mathcal{R}(r_s)}}}.
\end{align}

\section{Ray-tracing method}
\label{app:RayTracing}

A light ray shot back from Cartesian position $(\alpha,\beta)$ on the observer screen has radial trajectory (Eq.~(30) of Ref.~\cite{GrallaLupsasca2020a})
\begin{align}
	\label{eq:SourceRadius}
	r_s(I_r)=\frac{r_4r_{31}-r_3r_{41}\sn^2\pa{\frac{1}{2}\sqrt{r_{31}r_{42}}I_r-\mathcal{F}_o\big|k}}{r_{31}-r_{41}\sn^2\pa{\frac{1}{2}\sqrt{r_{31}r_{42}}I_r-\mathcal{F}_o\big|k}},
\end{align}
with $I_r$ denoting the Mino time elapsed along the radial trajectory from emission point to observer, and
\begin{align}
	\mathcal{F}_o=F\pa{\left.\arcsin{\sqrt{\frac{r_{31}}{r_{41}}}}\right|k},\quad
	k=\frac{r_{32}r_{41}}{r_{31}r_{42}}.
\end{align}
Here, we introduced the notation
\begin{align}
	r_{ij}=r_i-r_j,
\end{align}
with the roots $\cu{r_1,r_2,r_3,r_4}$ functions of $(\alpha,\beta,\theta_o)$ given in Ref.~\cite{GrallaLupsasca2020b}.  In practice, for each of the radial types of motion that we encounter \cite{GrallaLupsasca2020b}, we use an alternative (equivalent) form of Eq.~\eqref{eq:SourceRadius} that is manifestly real and therefore numerically more stable.

The maximal Mino time elapsed along the light ray before it returns to infinity (if outside the critical curve) or crosses the event horizon (if inside the critical curve) is
\begin{align}
	I_r^\mathrm{total}=
	\begin{cases} 
		\displaystyle2\int_{r_4}^{\infty}\frac{\ed r}{\sqrt{\mathcal{R}(r)}}&r_+<r_4\in\mathbb{R},\vspace{2pt}\\
		\displaystyle\int_{r_+}^{\infty}\frac{\ed r}{\sqrt{\mathcal{R}(r)}}&\text{otherwise},
	   \end{cases}
\end{align}
Closed-form expressions for this elliptic integral are given in App.~A of Ref.~\cite{GrallaLupsasca2020a}.  The Mino time elapsed in the polar trajectory from observer to $m^\text{th}$ equatorial crossing is given in Eq.~(20) of Ref.~\cite{GrallaLupsasca2020a} as
\begin{align}
	G_\theta^m&=\frac{1}{a\sqrt{-u_-}}\br{2mK\pa{\frac{u_+}{u_-}}-\sign\pa{\beta}F_o},\\
	F_o&=F\pa{\arcsin\pa{\frac{\cos{\theta_o}}{\sqrt{u_+}}}\left|\frac{u_+}{u_-}\right.},
\end{align}
where $u_\pm$ denote the roots of the angular potential, given in Eq~(11) therein.  Note that we have specialized to equatorial sources, which allows us to exclude vortical geodesics (with negative Carter constant) that can never reach the equator \cite{Kapec2020,GrallaLupsasca2020b}.

The Kerr geodesic equation $I_r=G_\theta$ (Eq.~(7a) of Ref.~\cite{GrallaLupsasca2020a}) requires the Mino times elapsed along the radial and polar motions to match.  Thus, $G_\theta$ is bounded above by $I_r^\mathrm{total}$ along the portion of the light ray in the Kerr exterior, but for each $m$ such that $G_\theta^m<I_r^\mathrm{total}$, the light ray crosses the equatorial plane at radius
\begin{align}
	\label{eq:EquatorialRadius}
	r_m(\alpha,\beta)=r_s(G_\theta^m)
\end{align}
outside the horizon.  The maximal value of equatorial crossings $m$ is thus
\begin{align}
	\label{eq:EquatorialCrossing}
	N(\alpha,\beta)=\left\lfloor\frac{a\sqrt{-u_-}+\sign\pa{\beta}F_o}{2K(u_+/u_-)}\right\rfloor.
\end{align}

\bibliography{moon}
\bibliographystyle{utphys}

\end{document}